\documentclass[%
aip,
floatfix,
amsmath,amssymb,
reprint%
]{revtex4-1}

\usepackage{graphicx}
\usepackage{dcolumn}
\usepackage{bm}

\usepackage[utf8]{inputenc}
\usepackage[T1]{fontenc}
\usepackage{mathptmx}
\usepackage{subfigure,xcolor}

\graphicspath{{./figures/}}
\usepackage[colorlinks=true,
citecolor=blue,
filecolor=black,	
linktoc=all,
linkcolor=black, 
urlcolor=red]{hyperref}

\begin{document}
	
\title[Depth and flatness]{The role of depth and flatness of a potential energy surface in chemical reaction dynamics.}
\author{Wenyang Lyu}
\author{Shibabrat Naik}
\email{s.naik@bristol.ac.uk}
\author{Stephen Wiggins}
\noaffiliation
\affiliation{School of Mathematics, University of Bristol\\ Fry Building, Woodland Road, Bristol BS8 1UG, United Kingdom}

\date{\today}

\begin{abstract}
In this study, we analyze how changes in the geometry of a potential energy surface in terms of depth and flatness can affect the reaction dynamics. We formulate depth and flatness in the context of one and two degree-of-freedom (DOF) Hamiltonian normal form for the saddle-node bifurcation and quantify their influence on chemical reaction dynamics~\cite{borondo1996,garciagarrido_tilting_2019}. In a recent work, \citeauthor{garciagarrido_tilting_2019}~\cite{garciagarrido_tilting_2019} illustrated how changing the well-depth of a potential energy surface (PES) can lead to a saddle-node bifurcation. They have shown how the geometry of cylindrical manifolds associated with the rank-1 saddle changes en route to the saddle-node bifurcation. Using the formulation presented here, we show how changes in the parameters of the potential energy control the depth and flatness and show their role in the quantitative measures of a chemical reaction. We quantify this role of the depth and flatness by calculating the ratio of the bottleneck-width and well-width, reaction probability (also known as transition fraction or population fraction), gap time (or first passage time) distribution, and directional flux through the dividing surface (DS) for small to high values of total energy. The results obtained for these quantitative measures are in agreement with the qualitative understanding of the reaction dynamics.
\end{abstract}
	
\maketitle
	
	

\section{Introduction}

The topography of a potential energy surface (PES) plays a fundamental role in determining reaction paths and reaction mechanisms~\cite{wales_energy_2004,steinfeld_chemical_1989,levine_molecular_2009}. For a given chemical reaction, the potential energy surface (PES) describes the variation of the electronic energy with the nuclear coordinates within the Born–Oppenheimer approximation~\cite{wales_energy_2004}. The electronic structure calculations generate a landscape of mountain ranges with peaks (local maxima) and valleys (local minima) with varying depth and flatness. One approach of crossing the mountain ranges is by going over the lowest point called the index-1 saddle~\cite{agaoglou_chemical_2019} and this mechanism is quite common in chemical reactions. This indicates that the potential energy difference between the saddle and the bottom of the valley, that is the \emph{depth}, and the gradient of the landscape, that is the \emph{flatness}, dictates the rate and volume of crossings of the saddle. Thus, the role of depth and flatness in crossing the saddle is relevant for understanding reaction dynamics. Furthermore, the asymmetry in the depth of a potential well on either side of an index-1 saddle can lead to difference in forward and backward reaction rates. In addition, it has been noted in \textit{ab initio} calculations that comparable flatness in different parts of a PES implies similar stability of isomers in those regions~\cite{preuss_theoretical_1979,bittererova_abinitio_1999}. Depth and flatness has also been linked with altering product ratios, delaying formation of specific isomers, difficulty in identifying the intrinsic reaction coordinate~\cite{koseki_intrinsic_1989,nummela_nonstatistical_2002}.  For example, the roaming phenomenon appears to be significantly influenced by a flat region of the potential energy surface resulting from long range interactions~\cite{bowman_roaming_2011,shepler_roaming_2011,mauguiere_roaming_2017} and potential energy surfaces describing reactions of organic molecules are characterized by post transition state bifurcations where a valley ridge inflection point is believed to be a significant geometrical feature~\cite{hare_post-transition_2017}. However, the relationship between depth or flatness and the quantitative measures of reaction dynamics has not been investigated in a way that can be connected with the qualitative understanding of the reactions. In this article, we present an approach for quantifying the role of depth and flatness in reaction dynamics by applying the proposed definition to a model Hamiltonian.

Since the concepts based on configuration space such as width of the bottleneck play an important role in rate calculations, a comparative study of the effects of the depth and flatness of a PES for these quantities is needed. We will present this comparison along side the phase space perspective which is the appropriate setting for the reaction dynamics. The phase space structures used in this study include the unstable periodic orbit associated with the index-1 saddle~\cite{agaoglou_chemical_2019} and its stable and unstable manifolds which have been discussed in a recent work~\cite{garciagarrido_tilting_2019}. These phase space structures explain the reaction mechanism that results from changing the depth and flatness of a PES. Thus, the geometry of the PES affects the dynamics which in turn affects the reaction rates.

This article is outlined as follows. In section~\ref{sect:models_methods}, we briefly describe the one and two DOF Hamiltonian normal form for the saddle-node bifurcation. Then, we give a formulation for the depth and flatness of a PES and apply the formulae to the one and two DOF systems along with visualizing the surfaces at different depth and flatness. In section~\ref{sect:results_discussion}, we discuss the influence of the depth and flatness of a PES on the ratio of the bottleneck-width and well-width, reaction probability, gap time distribution, and directional flux through the dividing surface. We present our conclusions and outlook in section IV.

\section{Models and Methods\label{sect:models_methods}}

\subsection{One degree-of-freedom saddle-node Hamiltonian}

The normal form for the one DOF Hamiltonian that undergoes a saddle-node bifurcation in phase space~\cite{Wiggins2017book} is given by
\begin{equation}
\mathcal{H}(x,p_x) = T(p_x) + V(x) = \frac{1}{2} \, p_x^2 - \sqrt{\mu} \, x^2 + \frac{\alpha}{3} \, x^3 \;,
\label{eqn:ham1dof_snbif}
\end{equation}
where parameter $\mu \geqslant 0$ controls the location of one of the equilibrium points relative to another and  $\alpha > 0$ is the well-depth parameter and denotes the strength of the nonlinear terms in the kinetic energy. The Hamiltonian vector field is given by
\begin{equation}
\begin{aligned}
\dot{x} & = \dfrac{\partial \mathcal{H}}{\partial p_x} =  p_x, \\
\dot{p_x} & =-\dfrac{\partial \mathcal{H}}{\partial x} =  2\sqrt{\mu} x - \alpha x^2.
\end{aligned}
\end{equation}
The two equilibrium points (also known as critical points of the PES) are located at $\mathbf{x}_1^e= (0,0)$ and $\mathbf{x}_2^e=(2\sqrt{\mu}/\alpha,0)$, and the energy of these equilibrium points are 
\begin{align}
\mathcal{H}(\mathbf{x}_1^e) = 0 \;,\quad \mathcal{H}(\mathbf{x}_2^e) = - \dfrac{4\mu^{3/2}}{3\alpha^2}    
\end{align}

Linear stability analysis~\cite{garciagarrido_tilting_2019} of these equilibrium points gives that $\mathbf{x}_1^e= (0,0)$ is a saddle and $\mathbf{x}_2^e=(2\sqrt{\mu}/\alpha,0)$ is a center equilibrium point. It is to be noted that the one DOF saddle-node Hamiltonian is integrable for all parameter values and trajectories lie on the isoenergetic contours given by the Hamiltonian~\eqref{eqn:ham1dof_snbif}. 

\begin{figure}[!ht]
	\centering
	\includegraphics[width=0.9\linewidth]{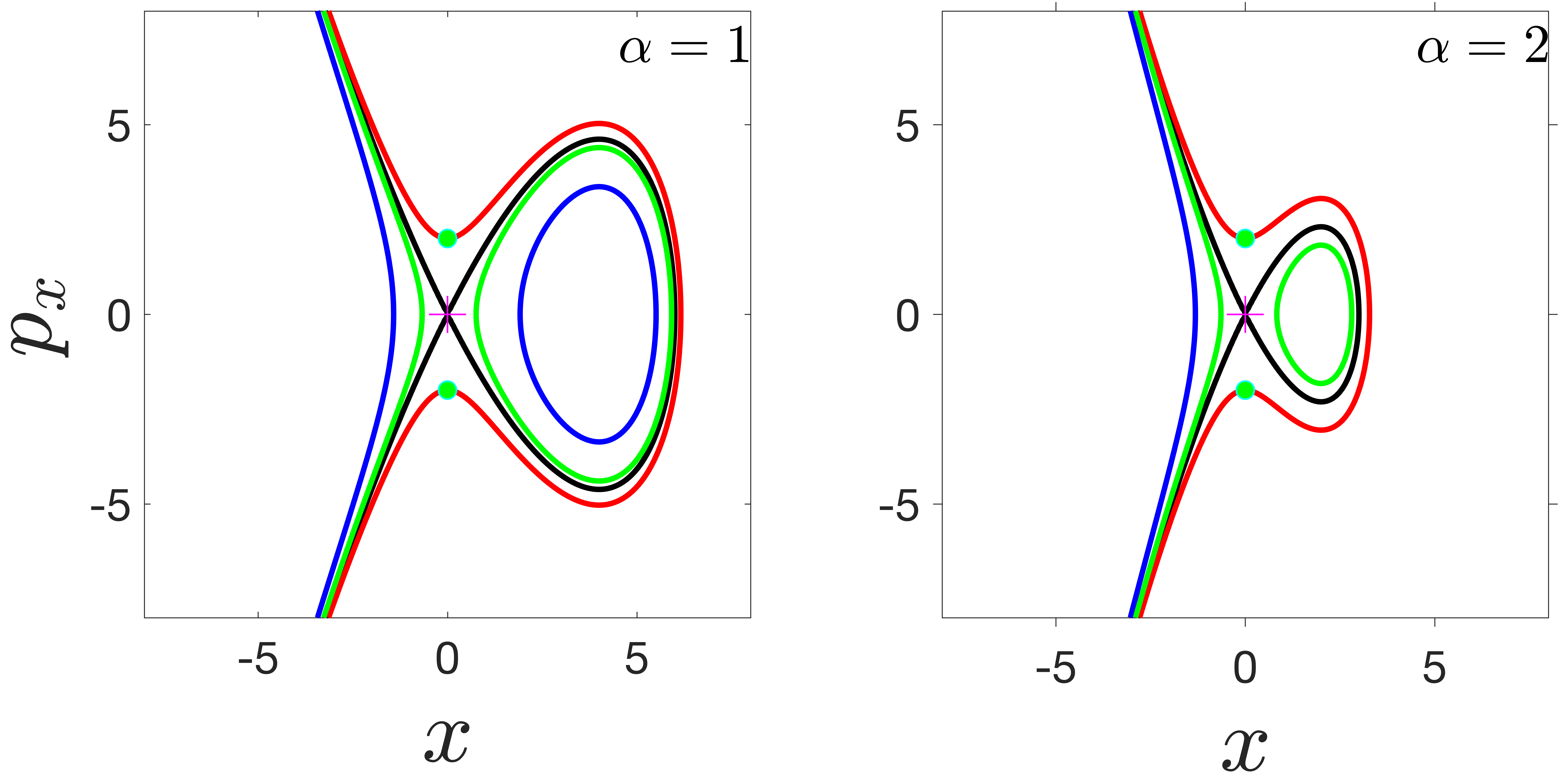}
	\caption{\textbf{Phase space of the one DOF saddle-node Hamiltonian} showing the change in the geometry of the reactive trajectories (shown in red, $e = 2$) and non-reactive trajectories (shown in blue $e = -5$ and green $e = -2$) for different values of the parameter $\alpha$ and fixed $\mu = 4$.}
	\label{fig:saddlenode1dof_mu4}
\end{figure}

In this system, we define the \emph{reaction} as the change in sign of the $x$-coordinate, and in particular, we specify reaction to be the event when a trajectory goes from $x > 0$ to $x < 0$. The geometry of the phase space structures can now be used to explain the mechanism behind the reactive and trapped trajectories~\cite{uzer2002geometry,wig2016}. The phase space structure in the bottleneck is the saddle equilibrium point at the origin which is a normally hyperbolic invariant manifold (NHIM)~\cite{wig2016,wiggins2013normally}. We note here that only for a one dimensional PES the NHIM (shown as red plus in Fig.~\ref{fig:saddlenode1dof_mu4}) does not change with total energy of the system, and in general the NHIM depends on the total energy. Next, the trajectories can be separated by constructing a dividing surface at total energy $\mathcal{H}(x,p_x) = e$. These are the points (shown as cyan dots) on the isoenergetic contour (shown as red curve) above the energy of the saddle equilibrium point in the Fig.~\ref{fig:saddlenode1dof_mu4}. The reaction dynamics at different total energies can now be classified as the reactive trajectories shown as red and black curves, or non-reactive trajectories shown as green or blue curves, respectively. As the parameters of the potential energy surface are varied, the geometry of the reactive trajectories in the phase space can be inferred from the isoenergetic contours as shown in the Fig.~\ref{fig:saddlenode1dof_mu4}. We also note that going from $\alpha = 1$ to $\alpha = 2$, the phase space volume inside the isoenergetic curves for $e \geqslant 0$ (bounded by the red curve and to the right of the origin) decreases. This implies that we need to enforce equal density when initializing reactant volume for calculations comparing different parameter values. We will return to this system after developing the formulation for depth and flatness. 

\subsection{Two degree-of-freedom saddle-node Hamiltonian}

In this section we introduce the normal form for the two DOF Hamiltonian that undergoes a saddle-node bifurcation by extending the one DOF Hamiltonian discussed above. To do so, we add another DOF in the form of a harmonic oscillator with mass $m = 1$ and frequency $\omega$. This new coordinate is referred to as a bath (or the perpendicular DOF) mode and may represent a vibrational DOF that does not break during the reaction. The influence of this DOF on the reaction coordinate can be parametrized by a quadratic coupling between the reaction and bath DOF. Thus, the Hamiltonian becomes 
\begin{align}
\mathcal{H}(x,y,p_x,p_y) & = T(p_x,p_y) + V(x,y) \nonumber \\
& = \dfrac{1}{2} \left(p_x^2 + p_y^2 \right) - \sqrt{\mu} \, x^2  + \frac{\alpha}{3} \,x^3 \notag \\
& \qquad \qquad \quad + \dfrac{\omega^2}{2} y^2 + \dfrac{\varepsilon}{2} \left(x-y\right)^2 \; ,
\label{eqn:ham_2dof}
\end{align}

where $\alpha > 0$ and $\mu \geq 0$ are the same as in the one DOF model system, $\omega > 0$ is the frequency of the harmonic oscillator or the bath mode, and $\varepsilon \geqslant 0$ is the coupling strength between the reaction and the bath mode. Identifying this Hamiltonian's kinetic energy, $T(p_x,p_y)$, and potential energy, $V(x,y)$, we get $T(p_x,p_y) = \frac{1}{2}\left(p_x^2 + p_y^2\right)$ and
\begin{equation}
V(x,y) = - \sqrt{\mu} \, x^2 + \frac{\alpha}{3} \,x^3 + \dfrac{\omega^2}{2} y^2 + \dfrac{\varepsilon}{2} \left(x-y\right)^2. \label{pes_2dof}
\end{equation}

The corresponding Hamilton's equations are given by:
\begin{equation}
\left.
\begin{aligned}
\dot{x} &= \dfrac{\partial \mathcal{H}}{\partial p_x} =  p_x \\
\dot{y} &= \dfrac{\partial \mathcal{H}}{\partial p_y} = p_y \\
\dot{p_x} &= -\dfrac{\partial \mathcal{H}}{\partial x} =  -\alpha \, x^2+ 2\sqrt{\mu} \, x + \varepsilon (y - x)\\
\dot{p_y} &= -\dfrac{\partial \mathcal{H}}{\partial y} = -\omega^2 y + \varepsilon (x-y)
\end{aligned}
\right.
\label{hameq_2dof}
\end{equation}
In this Hamiltonian system, the phase space is four dimensional and since energy is conserved, the trajectories evolve on a three dimensional energy surface. The equilibria for this system are located at $\mathbf{x}_1^e = (0,0,0,0)$ and $\mathbf{x}_2^e = \left(x^e,y^e,0,0\right)$
where
\begin{equation}
x^e = \frac{1}{\alpha }\left( 2 \sqrt{\mu}- \frac{\omega^2\varepsilon}{\omega^2 + \varepsilon} \right) \;,\quad y^e = x^e \left( \frac{\varepsilon}{\omega^2 +\varepsilon}  \right).
\label{eqn:cSpace_eqCoords}
\end{equation}

The total energy of the equilibrium points are
\begin{align}
\mathcal{H}(\mathbf{x}_1^e) = & 0 \\
\mathcal{H}(\mathbf{x}_2^e) = & (x^e)^2 \left( -\frac{1}{3}\sqrt{\mu} + \frac{\omega^2\varepsilon}{6(\omega^2 + \varepsilon)} \right) = - \dfrac{\alpha}{6}(x^e)^3.
\label{energy_eqpts}
\end{align}


\subsection{``Depth'' and ``Flatness'': A heuristic formulation\label{ssect:depth_flatness_formulas}} 

In this section, we present a definition for the depth and flatness of a PES and apply the formulae to the one and two degrees of freedom saddle-node Hamiltonian.

\paragraph*{\textbf{Depth.}} We define the depth, $\mathcal{D}$, of the PES as the difference between the potential energy of the saddle equilibrium point located in the bottleneck and the potential energy of the centre equilibrium point located in the bottom of the well. In case of a single well and bottleneck, this becomes
\begin{equation}	
\mathcal{D} = V(\mathbf{x}_{\rm sad}) - V(\mathbf{x}_{\rm cen})
\end{equation}
where $\mathbf{x}$ denotes the configuration coordinates, $\mathbf{x}_{\rm sad}$ and $\mathbf{x}_{\rm cen}$ are the configuration coordinates of the saddle and centre equilibria.

For the one and two DOF saddle-node Hamiltonians, this definition leads to expressions~\ref{eqn:depth_sn1dof} and~\ref{eqn:depth_sn2dof_coupled}. It is to be noted that when $\varepsilon = 0$ in the two DOF system, the depth expression becomes independent of the frequency, $\omega$, of the bath mode, with dependence only on $\mu$ and $\alpha$. We will revisit this observation while discussing the results.

\paragraph*{\textbf{Flatness.}} We define the flatness, $\mathcal{F}$, of the PES as the mean norm of the gradient of the potential energy over a bounded domain. Thus, the flatness is given by
\begin{equation}
\mathcal{F} = \overline{ \bigg|\bigg| \dfrac{\partial V(\mathbf{x})}{\partial \mathbf{x}} \bigg|\bigg|_2 } \, , \quad \mathbf{x} \in \Omega
\label{eqn:flatness}
\end{equation}
where  $\overline{||\cdot||_2}$ represents the average of the Euclidean-norm of the gradient of the potential energy function evaluated at discrete points in a bounded domain $\Omega$. The Euclidean norm of a vector $\mathbf{x}= (x_1, ..., x_n)$ is defined as the square root of the sum of squares of $x_i$. 

A few remarks are worth noting here. Firstly, our objective is to obtain a numerical representation for both depth and flatness such that a PES can be assigned a number or two different potential energy surfaces can be compared. Thus, we have adopted the above definitions after extensive numerical experiments with different formulations. Secondly, we know that the flatness and curvature as features of the potential energy landscape are related via the first and the second derivative of the potential energy function~\cite{wales_energy_2004}. Thus, the flatness as defined in Eqn.~\eqref{eqn:flatness} is closely related to the force experienced by the molecule/atom undergoing a reaction, and thus can be used in justifying the quantitative measures from a chemical intuition standpoint. Thirdly, the Euclidean norm of the gradient of the potential energy function is merely a starting point, and it remains to be checked if other norms of the gradient are better measures of flatness. Fourthly, the definition for depth can be extended to systems with multiple bottlenecks and wells by associating each well with a saddle point and calculating the depth of each well relative to that saddle point.

\subsubsection{Application: One degree-of-freedom saddle-node Hamiltonian\label{sssect:1dof_application}}

In the one DOF Hamiltonian, the potential energy function is given by
\begin{equation}
V(x) = -\sqrt{\mu}\ x^2 + \dfrac{\alpha}{3} \ x^3
\end{equation}
and thus the depth of the PES becomes
\begin{equation}
\mathcal{D} = V(0) - V\left( \dfrac{2 \sqrt{\mu}}{\alpha} \right) = \dfrac{4\mu^{3/2}}{3\alpha^2}
\label{eqn:depth_sn1dof}
\end{equation}

For the one DOF system, the flatness of the PES becomes 
\begin{equation}
\mathcal{F} = \overline{ \bigg|\bigg| {\dfrac{ d V(x)}{ d x} } \bigg|\bigg|_{2} } = \overline{ || { -2 \sqrt{\mu}\ x + \alpha \ x^2 } ||_{2} } 
\label{eqn:flatness_sn1dof}
\end{equation}

where $\overline{||\cdot||_2}$ is the same as the absolute value and we calculate the flatness over some domain $\Omega = [a,b]$ for some real values $a < b$.

The above calculation implies depth increases with increasing $\mu$ and decreases with increasing $\alpha$, while the flatness increases with increasing both $\mu$ and $\alpha$.  This is an indication that depth and flatness can not be independently varied. It has also been reported~\cite{garciagarrido_tilting_2019} that increasing $\alpha$ \textemdash~the leading order term that controls the effect of nonlinear terms in the potential energy function~\eqref{eqn:ham1dof_snbif} \textemdash~decreases depth. So, as far as this model Hamiltonian is concerned, the same parameter can be varied to change both depth and flatness.

\subsubsection{Application: Two degree-of-freedom saddle-node Hamiltonian\label{sssect:2dof_application}}

For the two DOF Hamiltonian, the potential energy is 
\begin{equation}
V(x,y) = \dfrac{\alpha}{3}\, x^3- \sqrt{\mu} \, x^2 +\dfrac{\omega^2}{2} y^2 + \dfrac{\varepsilon}{2} \left(x-y\right)^2 \label{eqn:pes_2dof}
\end{equation}
Applying the definition, the depth, $\mathcal{D}_{\varepsilon}$, of the PES is 
\begin{equation}
\mathcal{D}_{\varepsilon} = \frac{1}{6 \alpha^2}\left(2 \sqrt{\mu} - \frac{\omega^2 \varepsilon}{\omega^2 +\varepsilon}\right)^3 
\label{eqn:depth_sn2dof_coupled}
\end{equation}

which is the potential energy difference between the saddle-center and the centre-center equilibrium points. We use $\varepsilon$ in the subscript to distinguish between the two DOF uncoupled ($\varepsilon=0$) and coupled ($\varepsilon \neq 0$) systems. For $\varepsilon = 0$, the above expression simplifies to Eqn.~\ref{eqn:depth_sn1dof}, and we will refer to the expression as $\mathcal{D}_0$ for the two DOF uncoupled system.

For the two DOF system, the flatness of the PES is given by 
\begin{align}
\mathcal{F} = & \overline{ \bigg|\bigg|\dfrac{\partial V(\mathbf{x})}{\partial \mathbf{x}} \bigg|\bigg|_2 } = \overline{ \bigg|\bigg| \left( \dfrac{\partial V(x,y)}{\partial x},\dfrac{\partial V(x,y)}{\partial y} \right) \bigg|\bigg|_2 } \notag \\
= & \overline{ \bigg|\bigg| \left( \alpha \, x^2- 2\sqrt{\mu} \, x + \varepsilon (x - y), \omega^2 y - \varepsilon (x-y) \right) \bigg|\bigg|_2 }  
\label{eqn:flatness_sn2dof}
\end{align}
where $\overline{||\cdot||_2}$ is the Euclidean-norm of the gradient of the potential energy function. 
the gradient of the potential energy function is a two dimensional vector $\left( \dfrac{\partial  V(x,y)}{\partial x},\dfrac{\partial V(x,y)}{\partial y} \right) $ and we calculate the flatness of the PES over some bounded domain $\Omega = [a_1,b_1] \times [a_2,b_2]$ in two dimensions for some real values $a_i < b_i, i=1,2$.

\begin{figure}[!tb]
	\centering
	\includegraphics[width=0.49\linewidth]{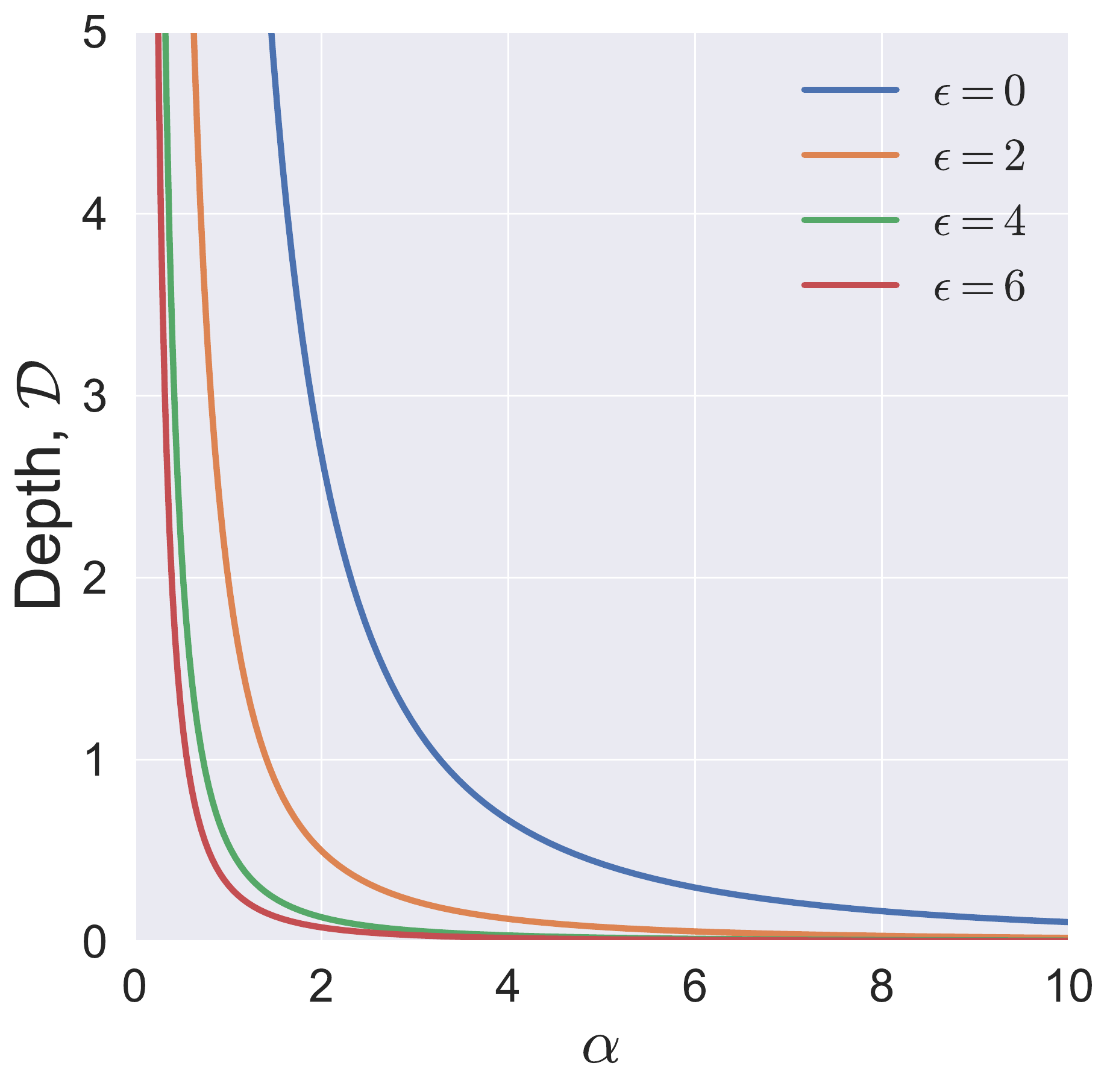}
	\includegraphics[width=0.49\linewidth]{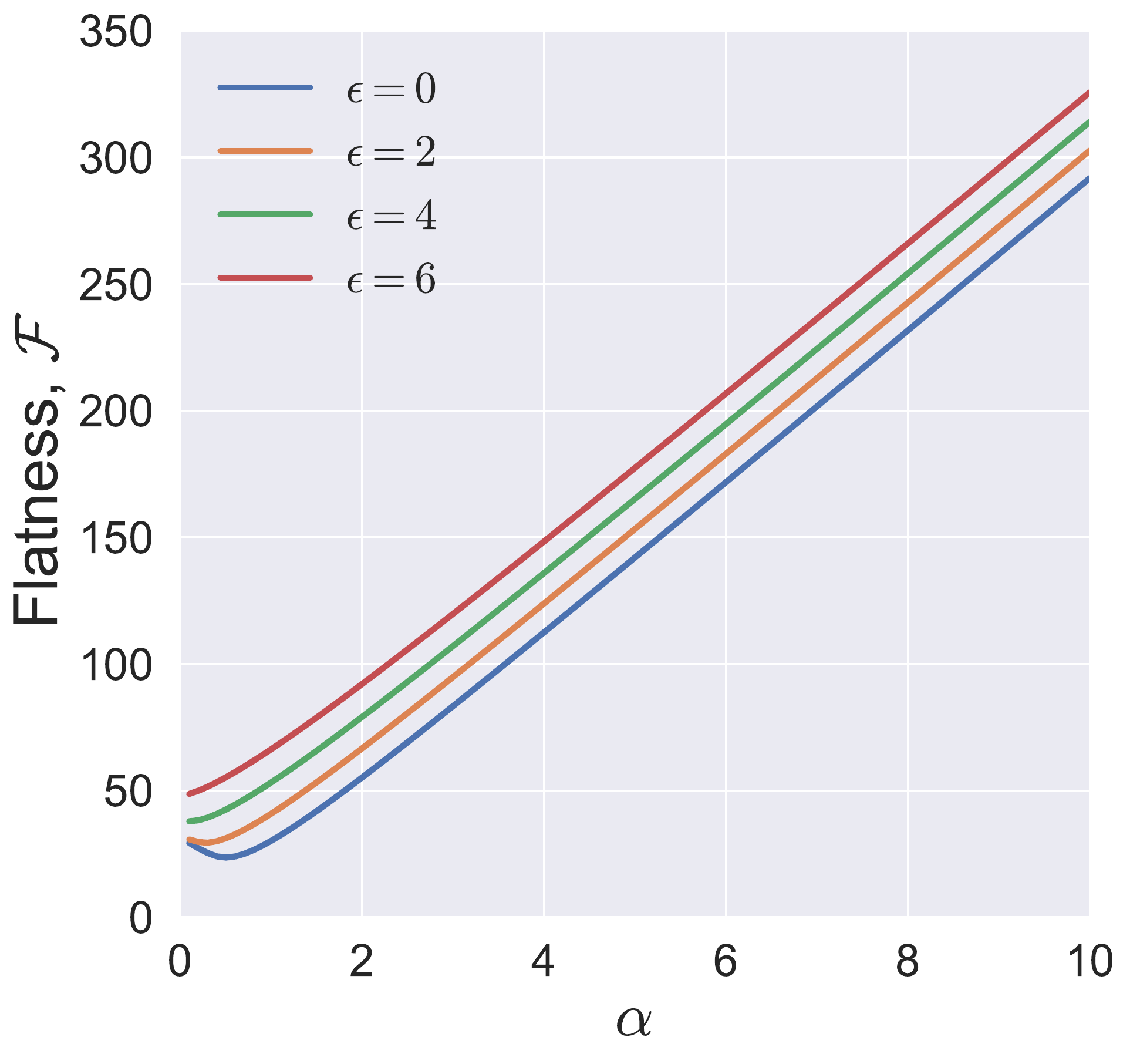}
	\caption{\textbf{Depth and flatness for the two DOF system} obtained by varying $\alpha$. Parameters are $\mu = 4,\omega = 3$ and the bounded domain $\Omega$ is $[-1,10] \times [3,3]$.}
\end{figure}

\begin{figure}[!tb]
	\centering
	\includegraphics[width=0.49\linewidth]{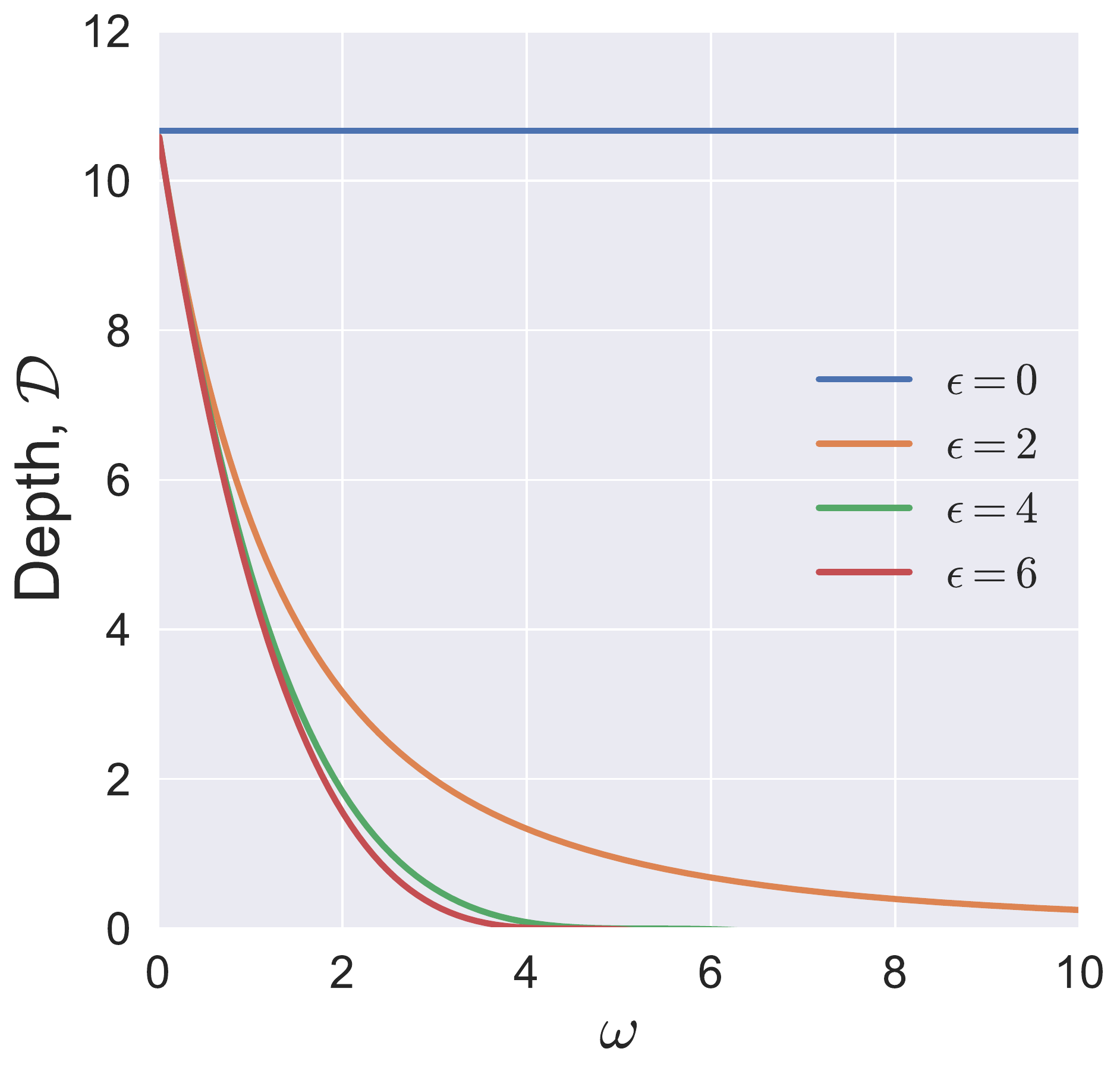}
	\includegraphics[width=0.49\linewidth]{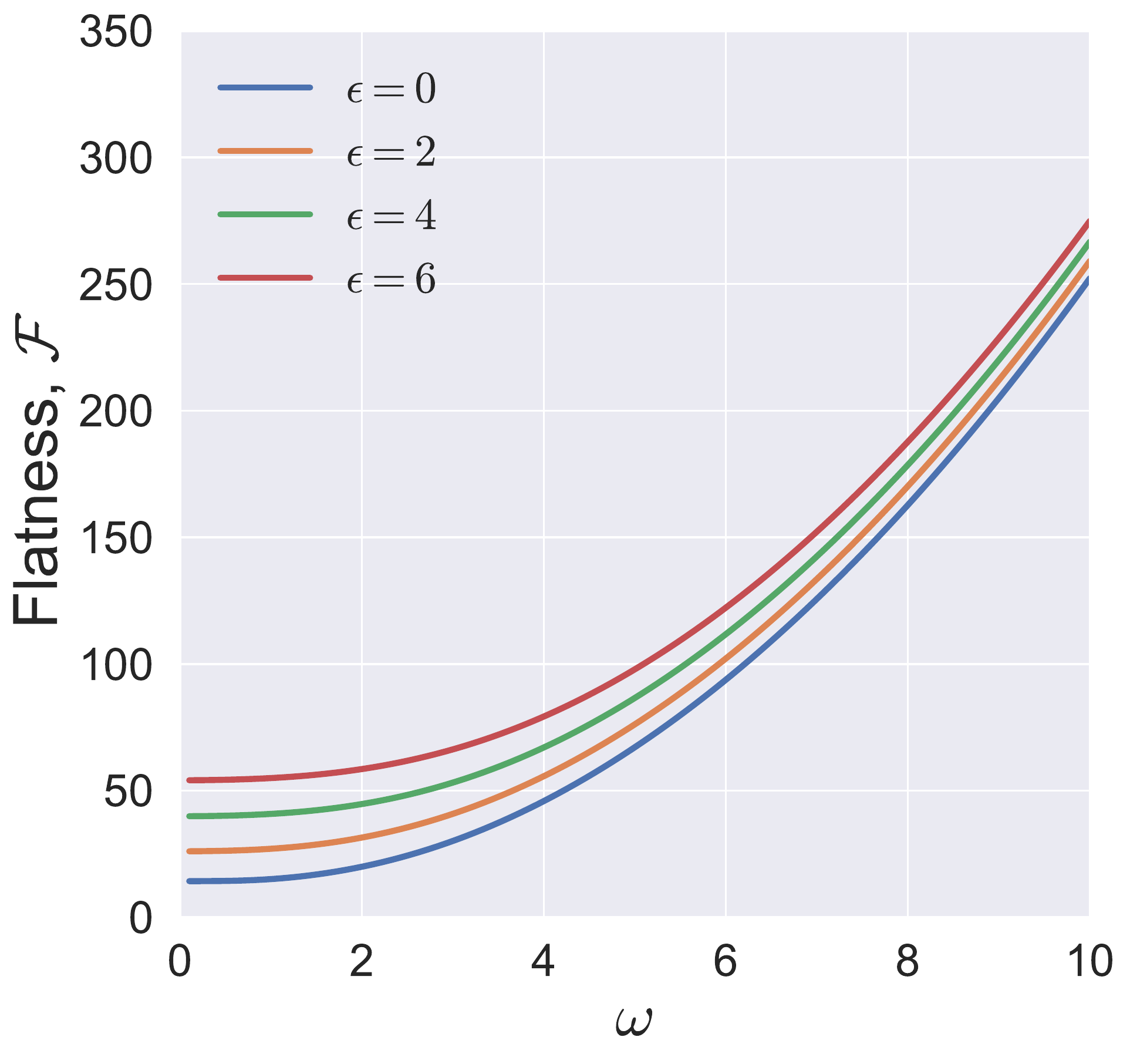}
	\caption{\textbf{Depth and flatness for the two DOF system} obtained by varying $\omega$. Parameters are $\mu = 4,\alpha = 1$ and the bounded domain $\Omega$ is $[-1,10] \times [3,3]$}
\end{figure}

\subsection{Visualizing potential energy surface with varying depth and flatness}

\begin{figure}[!ht]
	\includegraphics[width=0.45\textwidth]{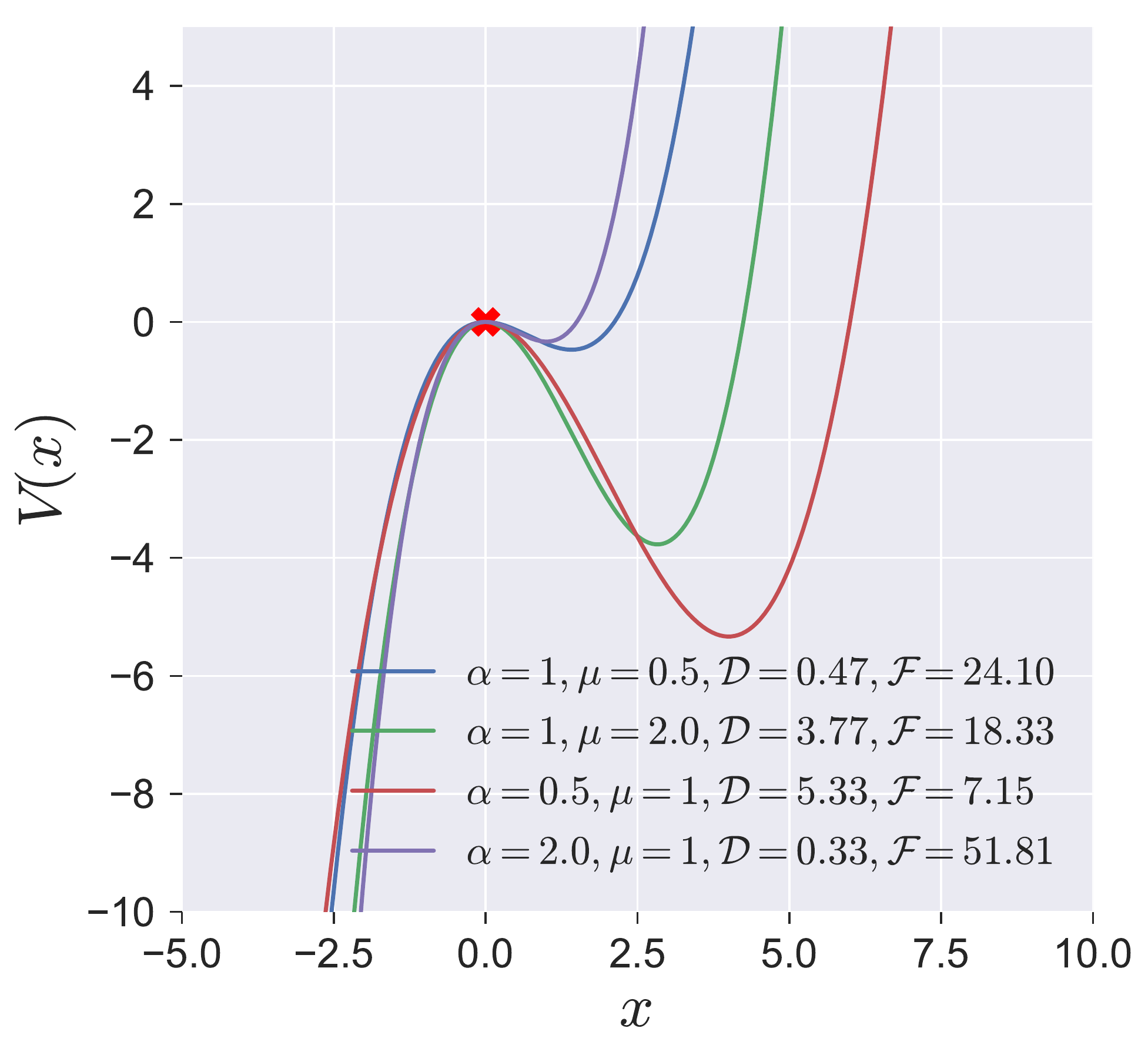}
	\caption{\textbf{Potential energy of the one DOF saddle-node Hamiltonian} for different depth and flatness obtained using the formula~\eqref{eqn:depth_sn1dof} and~\eqref{eqn:flatness_sn1dof}.}
	\label{fig:pes1dofalphamu}
\end{figure}

\begin{figure}
	\centering
	\includegraphics[width=0.49\linewidth]{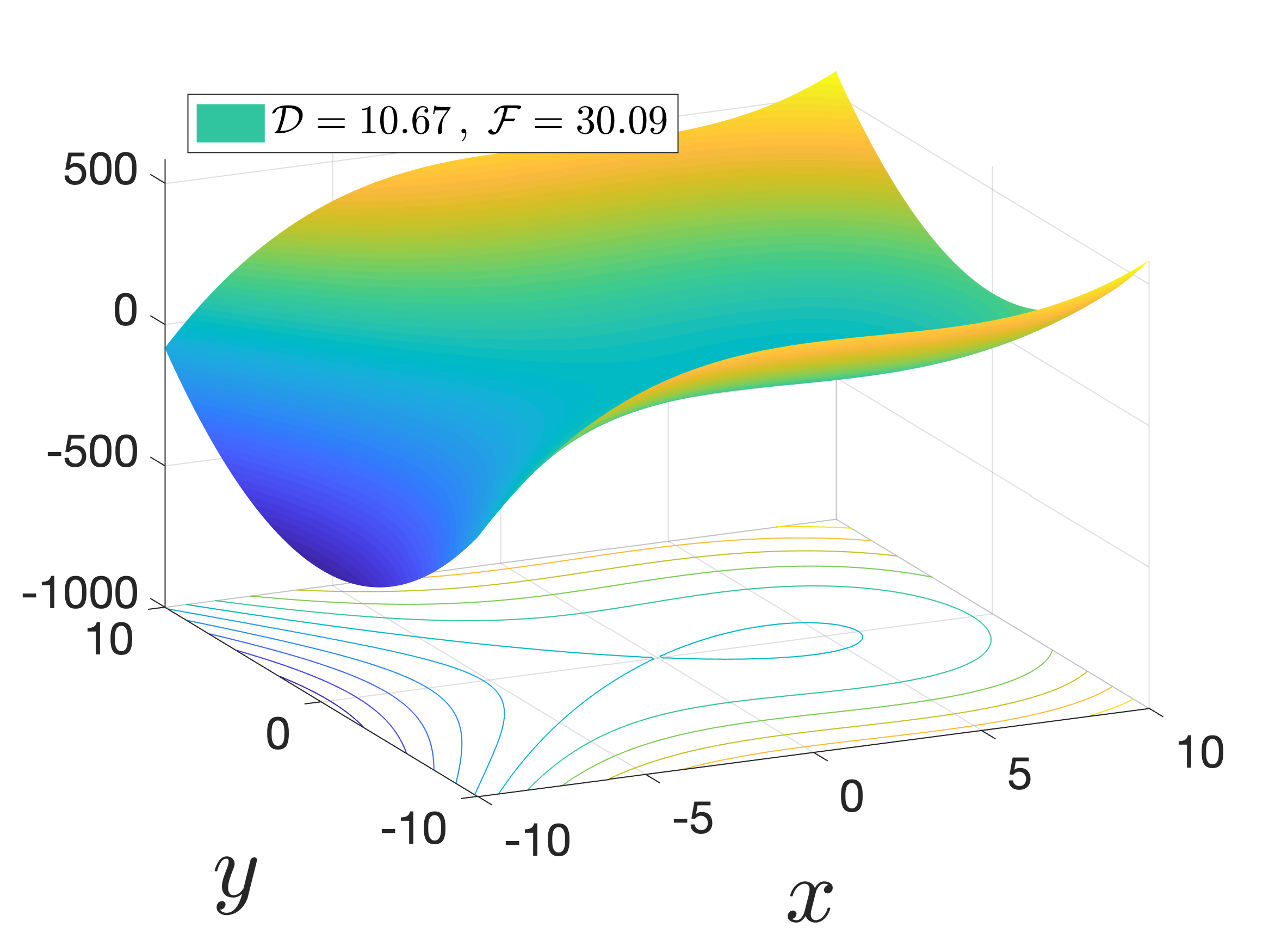}\label{fig:pes2dofmu4alpha1omega3epsilon0}
	\includegraphics[width=0.49\linewidth]{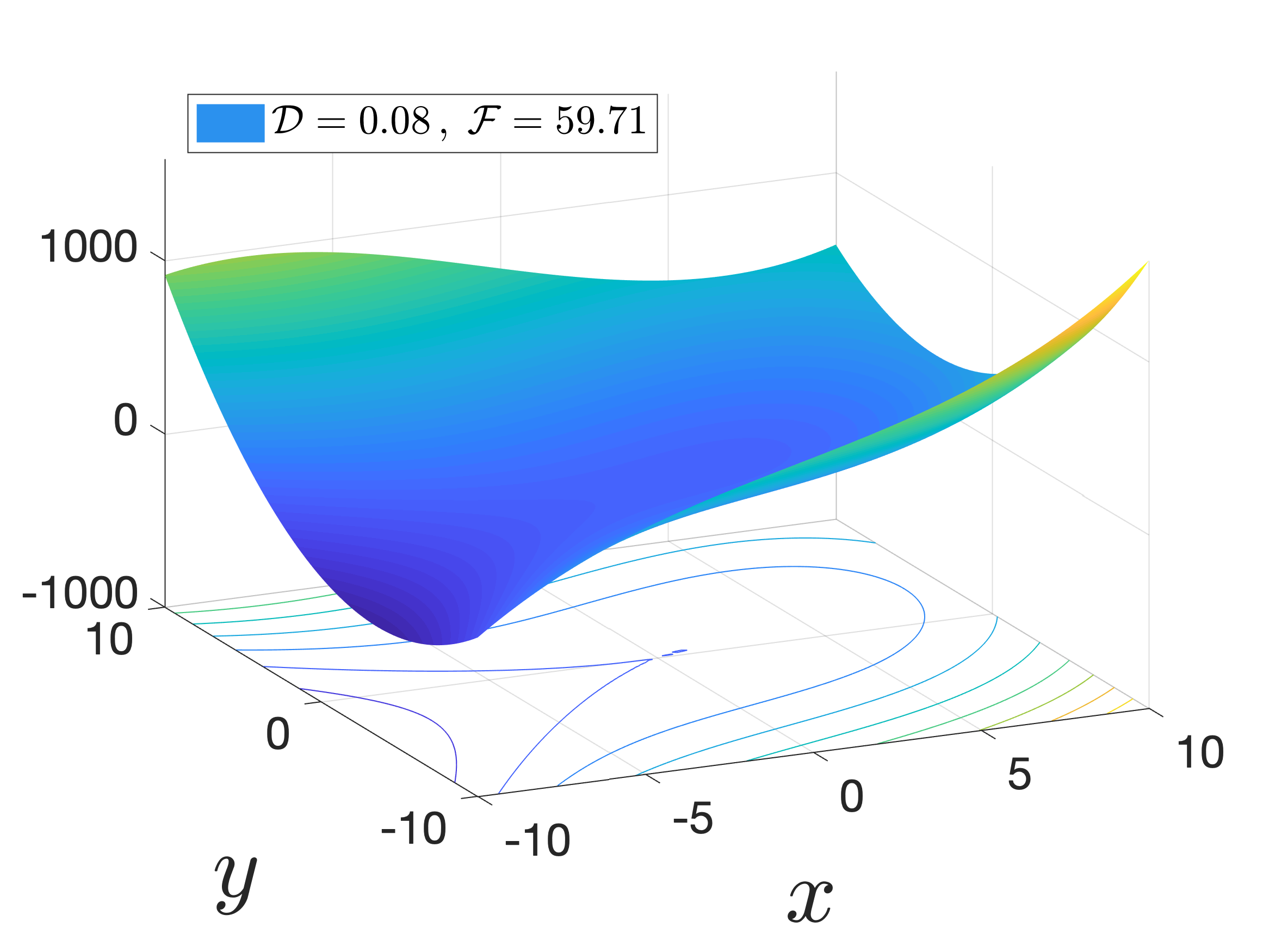}\label{fig:pes2dofmu4alpha1omega3epsilon5}
	\includegraphics[width=0.49\linewidth]{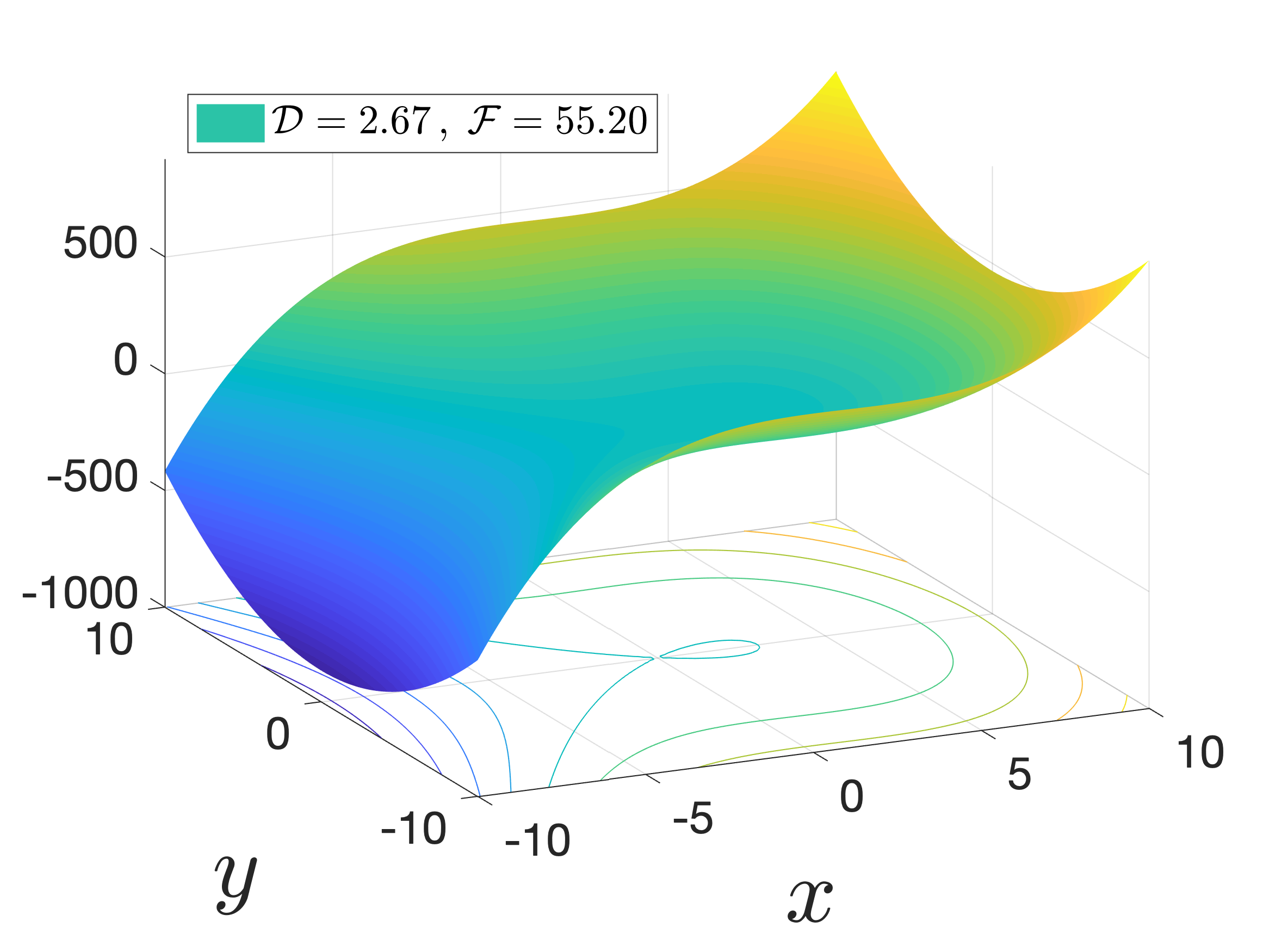}\label{fig:pes2dofmu4alpha2omega3epsilon0}
	\includegraphics[width=0.49\linewidth]{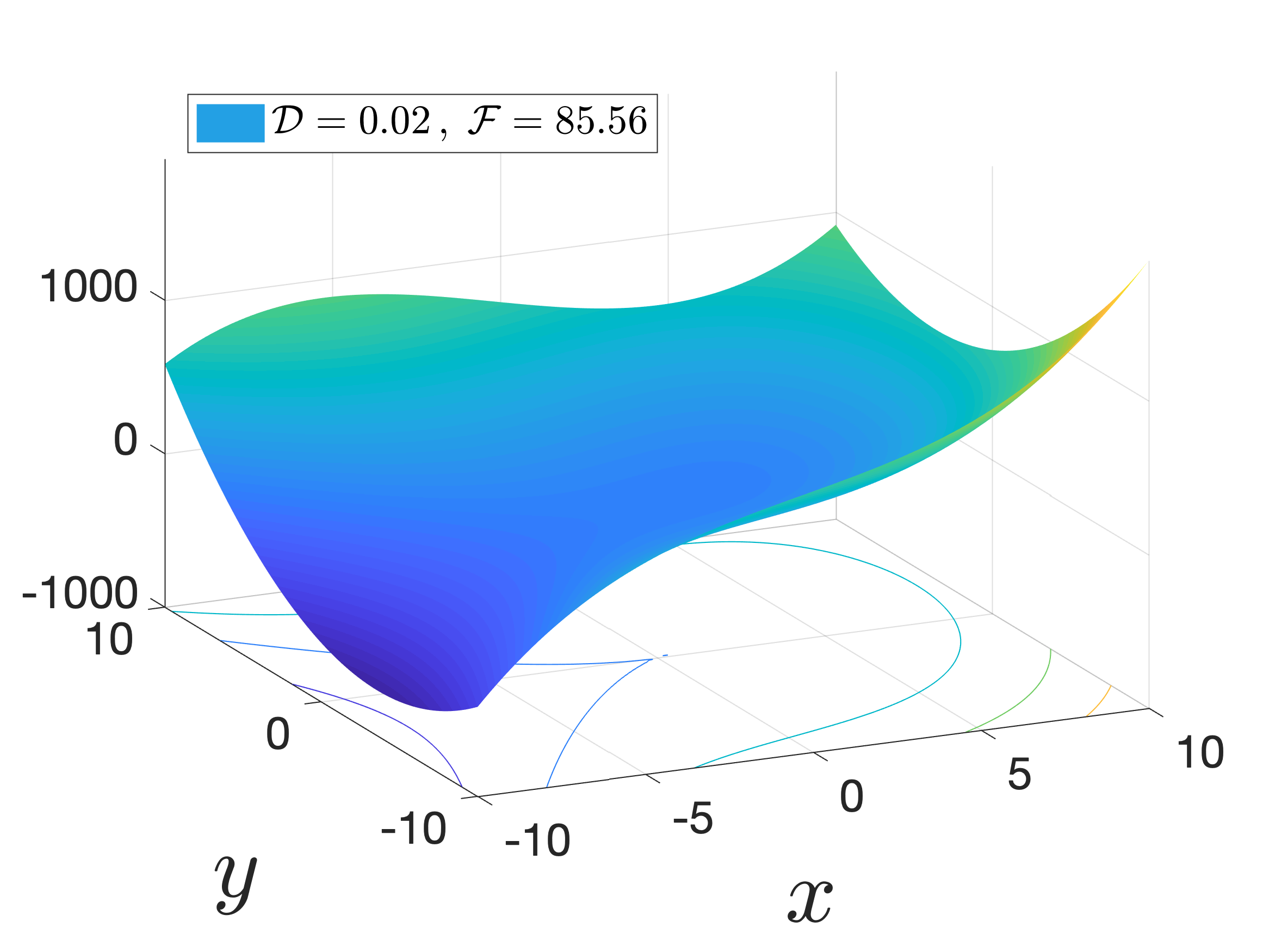}\label{fig:pes2dofmu4alpha2omega3epsilon5}
	\includegraphics[width=0.49\linewidth]{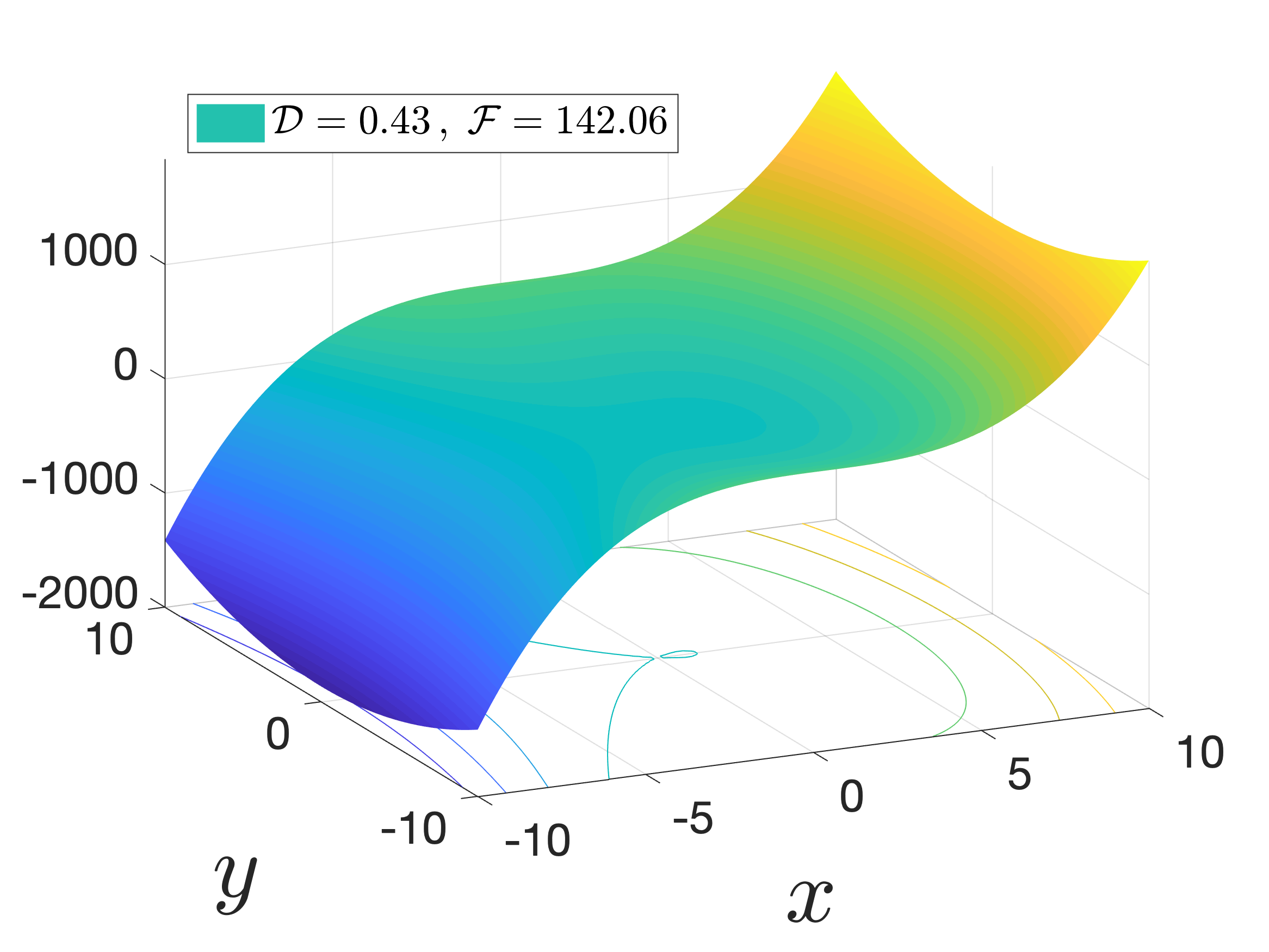}\label{fig:pes2dofmu4alpha5omega3epsilon0}
	\includegraphics[width=0.49\linewidth]{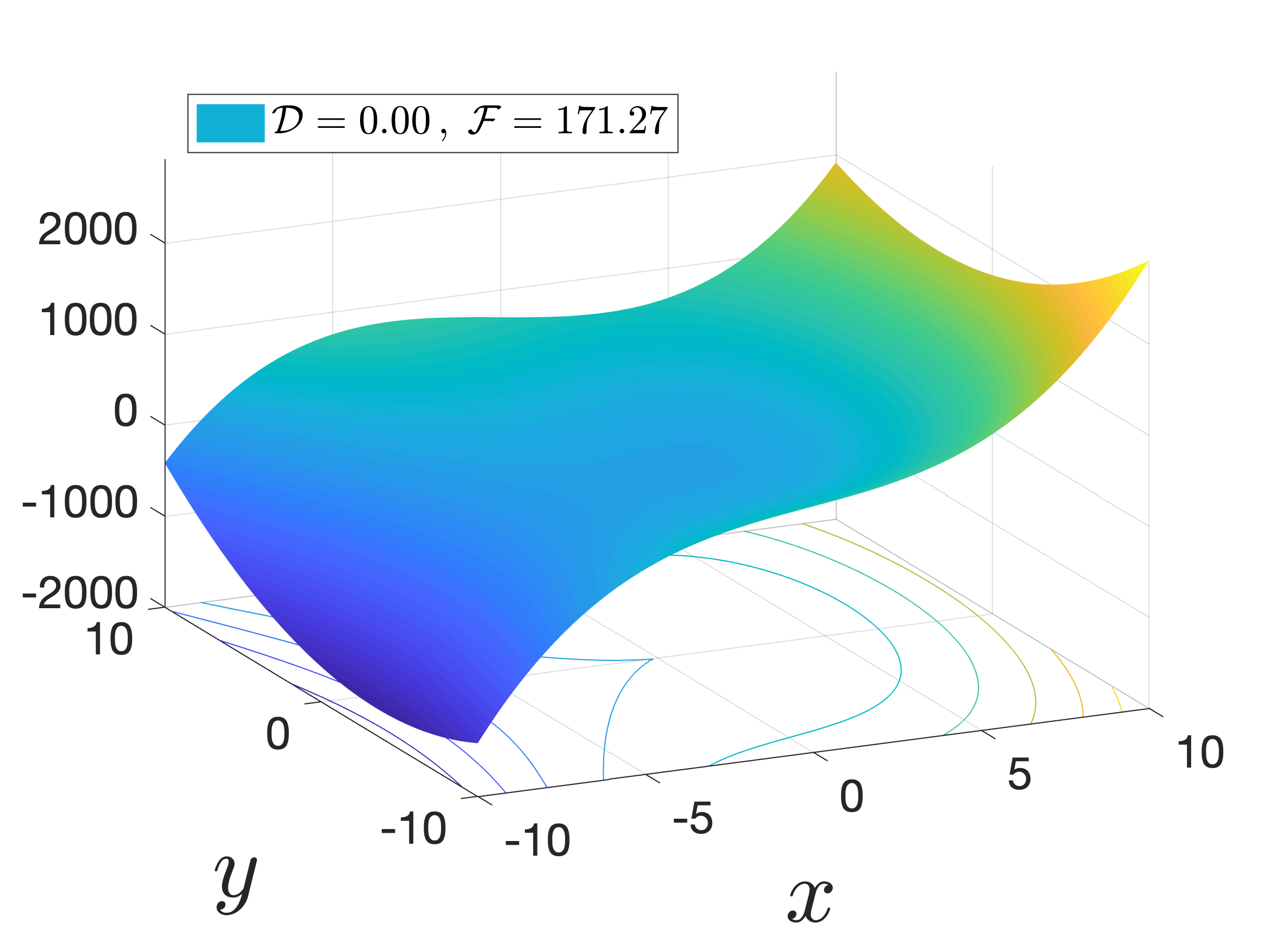}\label{fig:pes2dofmu4alpha5omega3epsilon5}
	\caption{\textbf{Potential energy surface of the two DOF saddle-node Hamiltonian} with different depth and flatness obtained using the formula~\eqref{eqn:depth_sn2dof_coupled} and~\eqref{eqn:flatness_sn2dof}. Left column shows the uncoupled system ($\varepsilon = 0.0$), right column shows the coupled system ($\varepsilon = 5.0$), and $\alpha = 1, 2, 5$ along the top, middle, and bottom row, respectively. In all the plots, other parameters are fixed: $\mu = 4, \omega = 3$.}
	\label{fig:pes2dof_alpha_epsilon}
\end{figure}

In this section, we visualize the qualitative changes in the potential energy function for different values of the depth and flatness in Fig.~\ref{fig:pes1dofalphamu} and~\ref{fig:pes2dof_alpha_epsilon}. In these plots, formulae derived in section: \ref{sssect:1dof_application} and \ref{sssect:2dof_application} are used to calculate the values shown. 

We observe that the depth and flatness are related in a way that increasing depth (cf. $\mu = 0.5$ and $\mu = 2.0$ for $\alpha = 1.0$ in Fig.~\ref{fig:pes1dofalphamu}) implies flatness will decrease, and the vice versa also holds (cf. $\alpha = 0.5$ and $\alpha = 2.0$ for $\mu = 1.0$ in Fig.~\ref{fig:pes1dofalphamu}). This inverse relationship between depth and flatness is also observed in the two degree-of-freedom Hamiltonian (Eqn.~\ref{eqn:ham_2dof}) for both the uncoupled and coupled cases as shown in Fig.~\ref{fig:pes2dof_alpha_epsilon}.

\section{Results and discussion\label{sect:results_discussion}} 

In this section, we use the depth and flatness formulations developed above to quantify their influence on the width of the bottleneck (to be referred to as \emph{bottleneck-width}) and the width of the potential well (to be referred to as \emph{well-width}), reaction probability, gap time distribution and directional flux through the DS. These quantities characterize reaction dynamics by capturing the changes in the reactive trajectory behavior which is being influenced by the changing of the depth and flatness of the PES.

\subsection{Ratio of the bottleneck-width and well-width}
In this subsection, we show the effect of the depth and flatness on the ratio of the bottleneck-width and well-width since this has been used as a measure of the changes in the shape of the PES when varying energy or other system parameters~\cite{de_leon_intramolecular_1981}. It is to be noted that for the two DOF system, we define this ratio in the configuration space which is two-dimensional. However, for the one DOF system, it is invalid to define this ratio in the one-dimensional configuration space, hence we define this ratio in the phase space, and not the one-dimensional configuration space.


\subsubsection{One degree-of-freedom Hamiltonian}

The bottleneck is defined at $x = 0$, for a given energy $\mathcal{H}(x,y,p_x,p_y)=e> \mathcal{H}(\mathbf{x}_1^e)$ where $\mathcal{H}(\mathbf{x}_1^e)=0$ is the total energy of the saddle equilibrium point. The width of the bottleneck, $w_b$ is defined as the difference of the $p_x$ coordinates between the two points on $\mathcal{H}(x,y,p_x,p_y)=e$ with $x=0$ which equals $ 2 \sqrt{2e}$. The width of the well, $w_w$ is defined as the difference of $p_x$ coordinates between the two points on $\mathcal{H}(x,y,p_x,p_y)=e$ with $x=2\sqrt{\mu}/\alpha$ which equals $2 \sqrt{2\left(e+ \dfrac{4\mu^{3/2}}{3\alpha^2}\right) }$. Thus, the ratio of the bottleneck-width and well-width, denoted by $R_{bw}$ becomes

\begin{align}
R_{bw} & = \dfrac{\sqrt{2e}}{\sqrt{2\left(e+ \dfrac{4\mu^{3/2}}{3\alpha^2}\right) }} = \sqrt{\dfrac{e}{e + \dfrac{4\mu^{3/2}}{3 \alpha^2}}} = \sqrt{\dfrac{e}{e + \mathcal{D}}}.
\end{align}

\begin{figure}[!ht]
	\centering
	\subfigure[]{\includegraphics[width=0.49\linewidth]{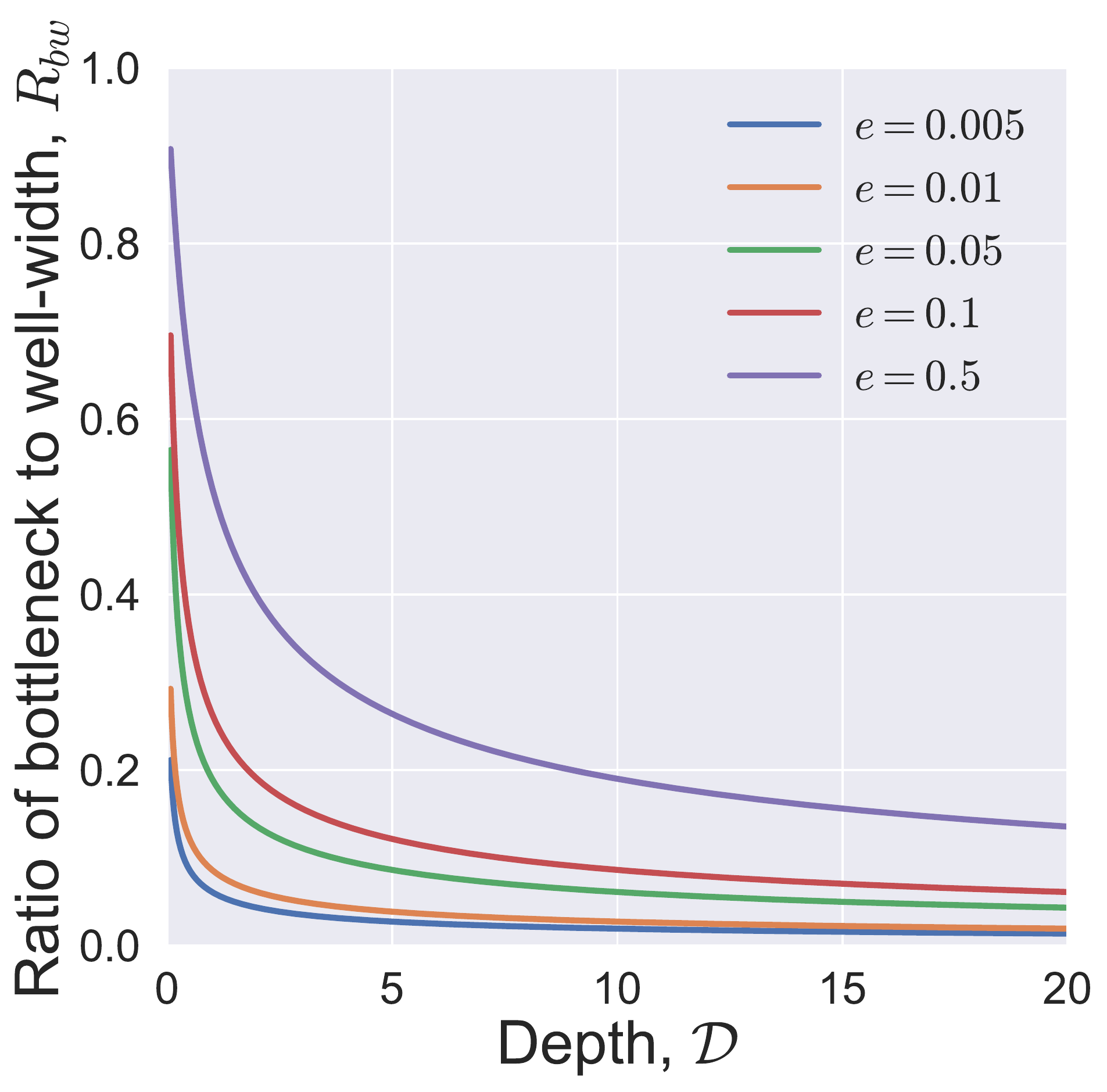}\label{fig:Rbw_1dof_dep_mu=4}}
	\subfigure[]{\includegraphics[width=0.49\linewidth]{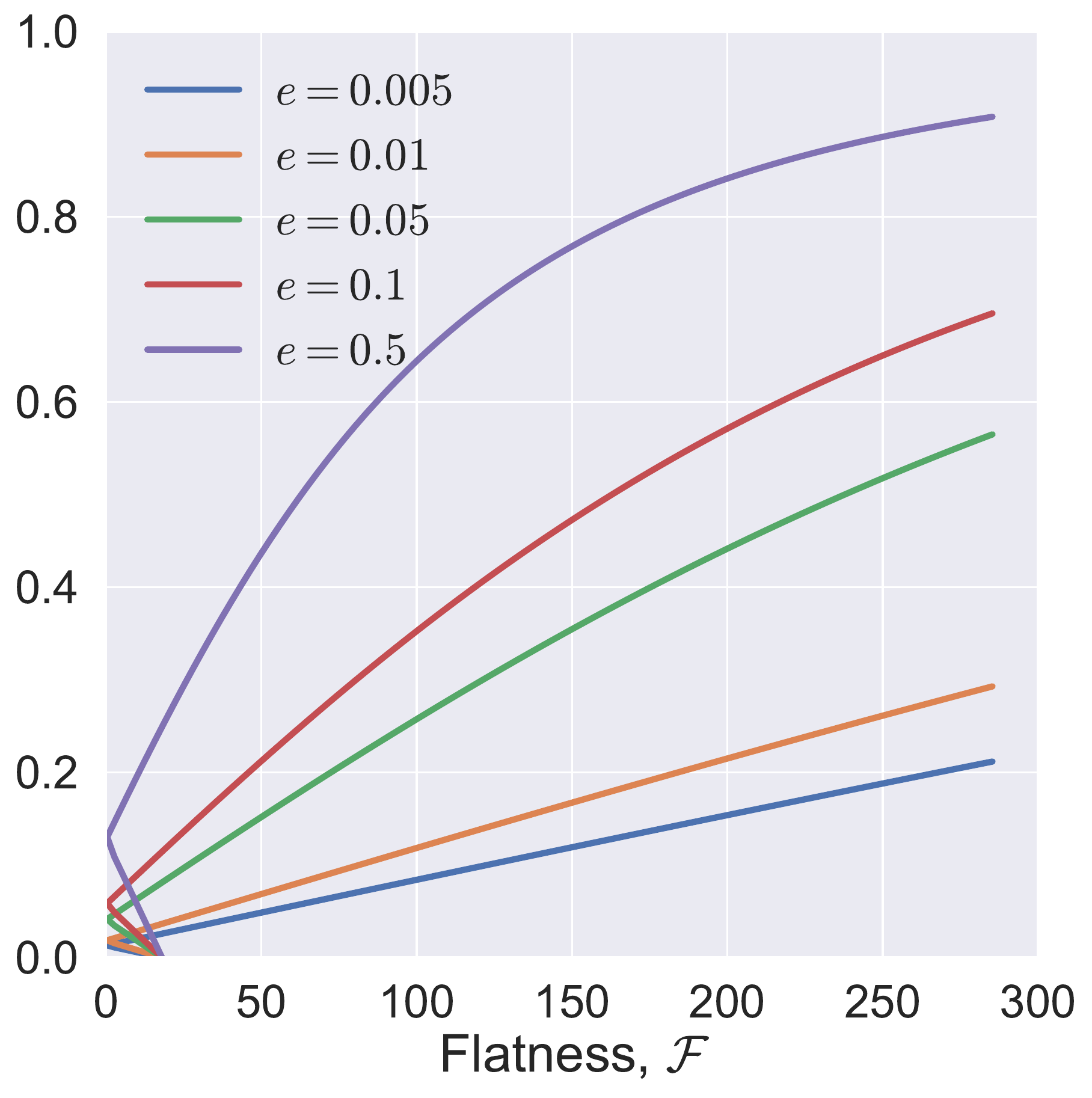}\label{fig:Rbw_1dof_flat_mu=4}}
	\caption{\textbf{Ratio of the bottleneck-width and well-width for the one DOF system.} Shows that the ratio decreases with increase in \protect\subref{fig:Rbw_1dof_dep_mu=4} depth, $\mathcal{D}$ and increases with increase in \protect\subref{fig:Rbw_1dof_flat_mu=4} flatness, $\mathcal{F}$. Depth and flatness is varied using the $\alpha$ parameter, while $\mu = 4$ is fixed in the one DOF model~\eqref{eqn:ham1dof_snbif}.}
\end{figure}
In the above derivation, we have subsituted the expression~\eqref{eqn:depth_sn1dof} to identify the relationship between the ratio, $R_{bw}$, and the depth of the PES and this relationship for different total energies is summarized in the Fig.~\ref{fig:Rbw_1dof_dep_mu=4}. To show the influence of the flatness on this ratio, we evaluate the expression~\eqref{eqn:flatness_sn1dof} numerically for different values of the system parameters and show the changes in terms of the flatness in Fig.~\ref{fig:Rbw_1dof_flat_mu=4}. Except for small values of the flatness, we can deduce that the ration increases with increasing flatness, where the monotonic increase becomes nonlinear for higher values of the total energy.

%
\subsubsection{Two degree-of-freedom Hamiltonian}

\begin{figure}[!ht]
	\centering
	\subfigure[]{\includegraphics[width=0.23\textwidth]{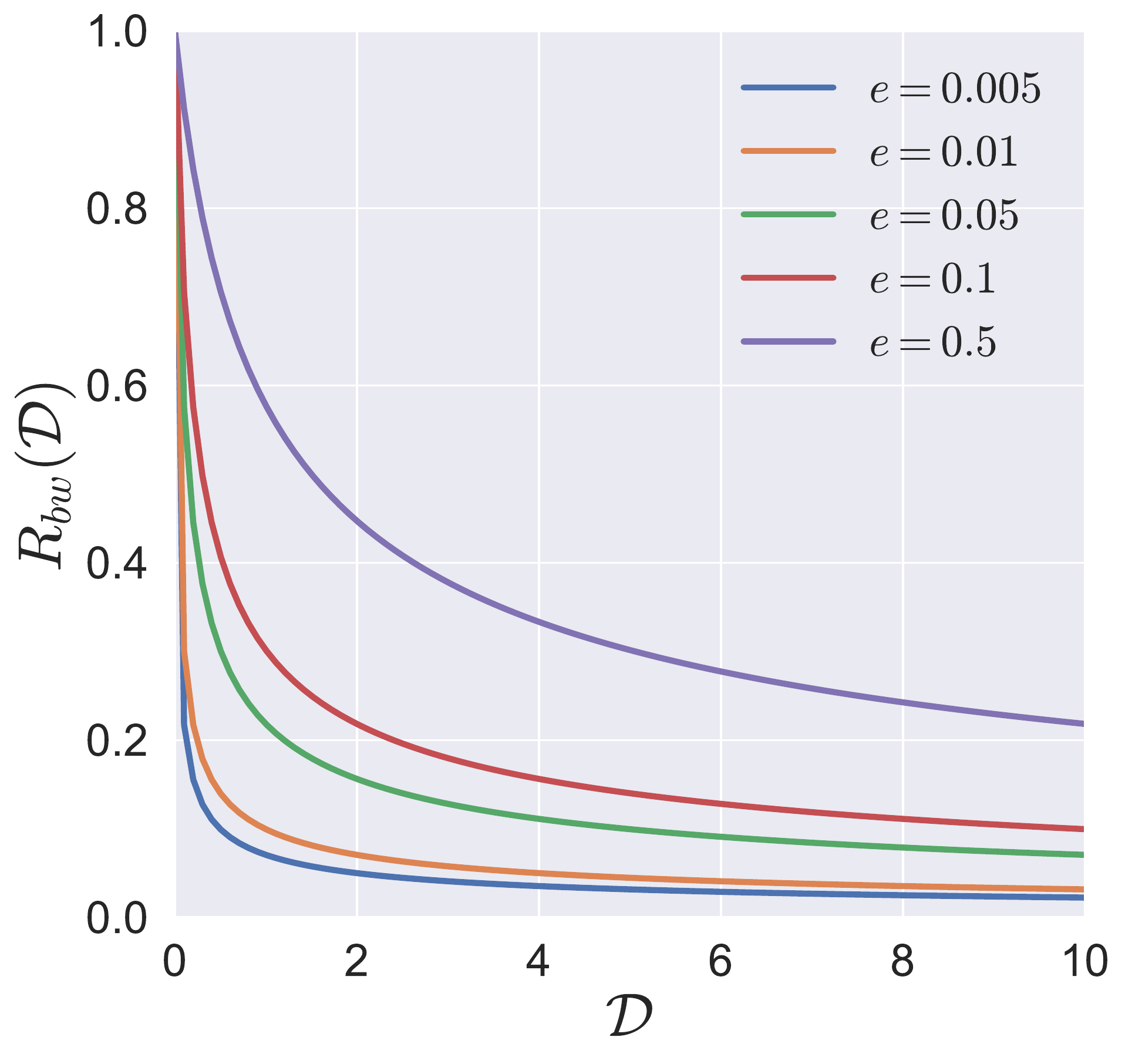}\label{fig:Rbw_2dof_uncoupled_depth}}
	\subfigure[]{\includegraphics[width=0.23\textwidth]{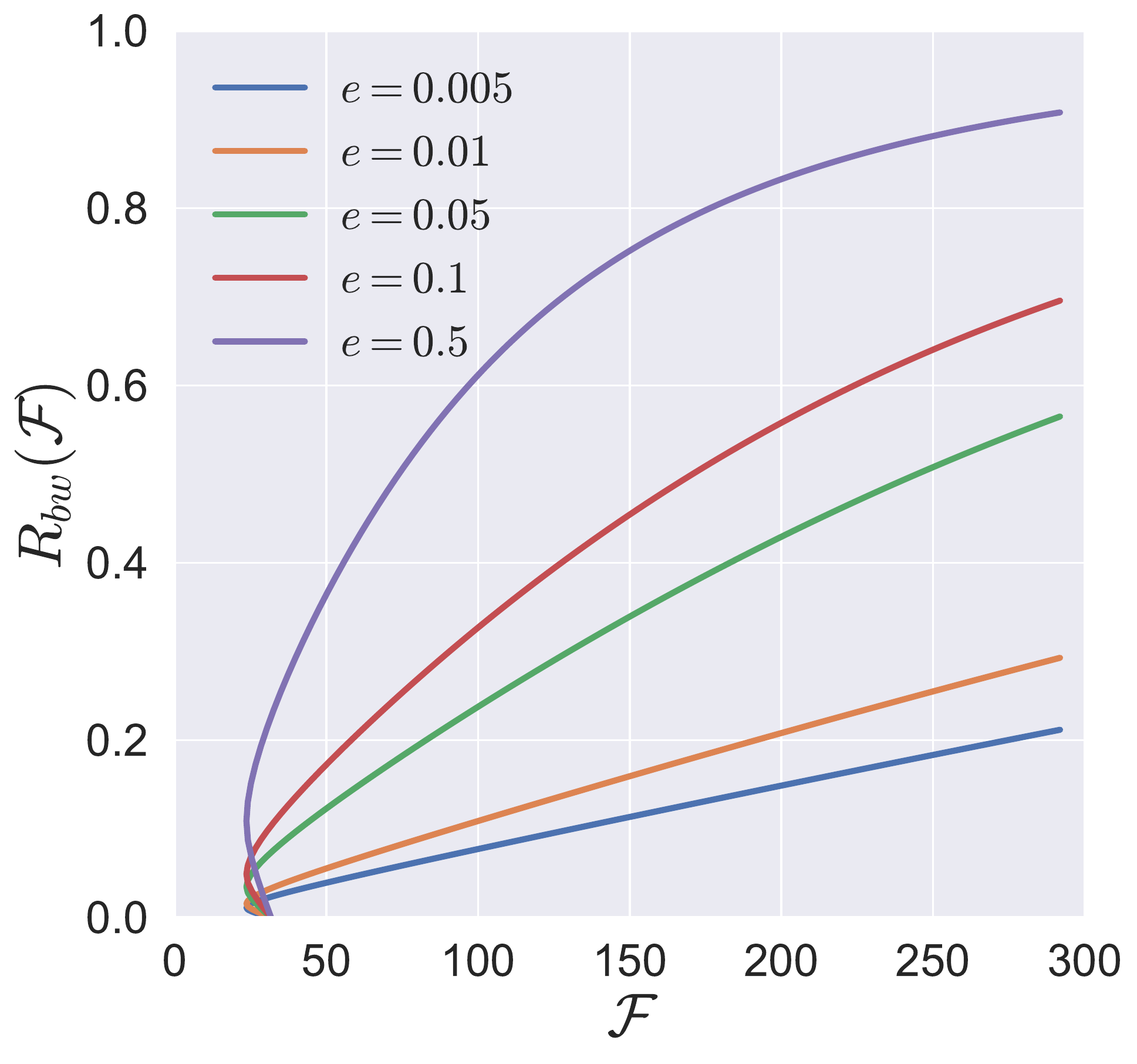}\label{fig:Rbw_2dof_uncoupled_flatness}}
	\caption{\textbf{Ratio of the bottleneck-width and well-width for the two DOF uncoupled system.} Shows the changes in the ratio for increasing (a) depth and (b) flatness of the PES at different total energies. System parameters $\mu=4,\omega=3,\varepsilon=0$ are fixed, while the depth and flatness of the PES is changed by varying the parameter $\alpha$.}
	\label{fig:Rbw_2dof_uncoupled}
\end{figure}

For the two DOF uncoupled system, that is $\varepsilon=0$, the bottleneck is open and can only be defined at $x=0$,  when $ e > \mathcal{H}(\mathbf{x}_1^e)$ where $\mathcal{H}(\mathbf{x}_1^e)=0$ is the total energy of the saddle equilibrium point.
The bottleneck-width in the configuration space, $w_b$ is defined as the difference in the $y-$coordinates between the two points on the PES, $V(x,y) = e$ at $x=0$, which equals $2 \sqrt{(2e)/\omega^2}$. We note that this definition uses the dynamical concept that in the bottleneck the maximum value of the potential energy $V(x,y)$ is achieved when the kinetic energy vanishes, that is $T(p_x,p_y) = 0$, which are the two turning points on the equipotential line, $V(x,y)=e$, in the configuration space. The width of the well, $w_w$, is defined as the difference of the $y-$coordinates between the two points on the PES, $V(x,y) = e$ where $x = x^e$ as in the Eqn.~\eqref{eqn:cSpace_eqCoords}, which equals $ 2 \sqrt{(2/\omega^2) \left( e + (4\mu^{3/2})/(3 \alpha^2) \right) }$. Thus, the ratio of the bottleneck-width and well-width, $R_{bw}$, becomes
\begin{equation}
R_{bw} = \dfrac{\sqrt{\dfrac{2e}{\omega^2}}}{\sqrt{\dfrac{2}{\omega^2} \left( e + \dfrac{4\mu^{3/2}}{3 \alpha^2} \right) }} = \sqrt{\dfrac{e}{e + \dfrac{4\mu^{3/2}}{3 \alpha^2}}} = \sqrt{\dfrac{e}{e + \mathcal{D}_{0}}}
\end{equation}

which is independent of $\omega$, the ``bath'' coordinate parameter, and where we have again substituted Eqn.~\eqref{eqn:depth_sn1dof} to simplify the dependence of the ratio on the depth. This relationship of the ratio, $R_{bw}$, and depth is shown in the Fig.~\ref{fig:Rbw_2dof_uncoupled_depth}.  Similar to the approach in the one DOF system, we show the influence of the flatness on this ratio by evaluating the expression~\eqref{eqn:flatness_sn2dof} ($\epsilon = 0$) numerically for different values of the system parameters and show the changes in the Fig.~\ref{fig:Rbw_2dof_uncoupled_flatness}. 


\begin{figure}[!ht]
	\centering
	\subfigure[]{\includegraphics[width=0.48\textwidth]{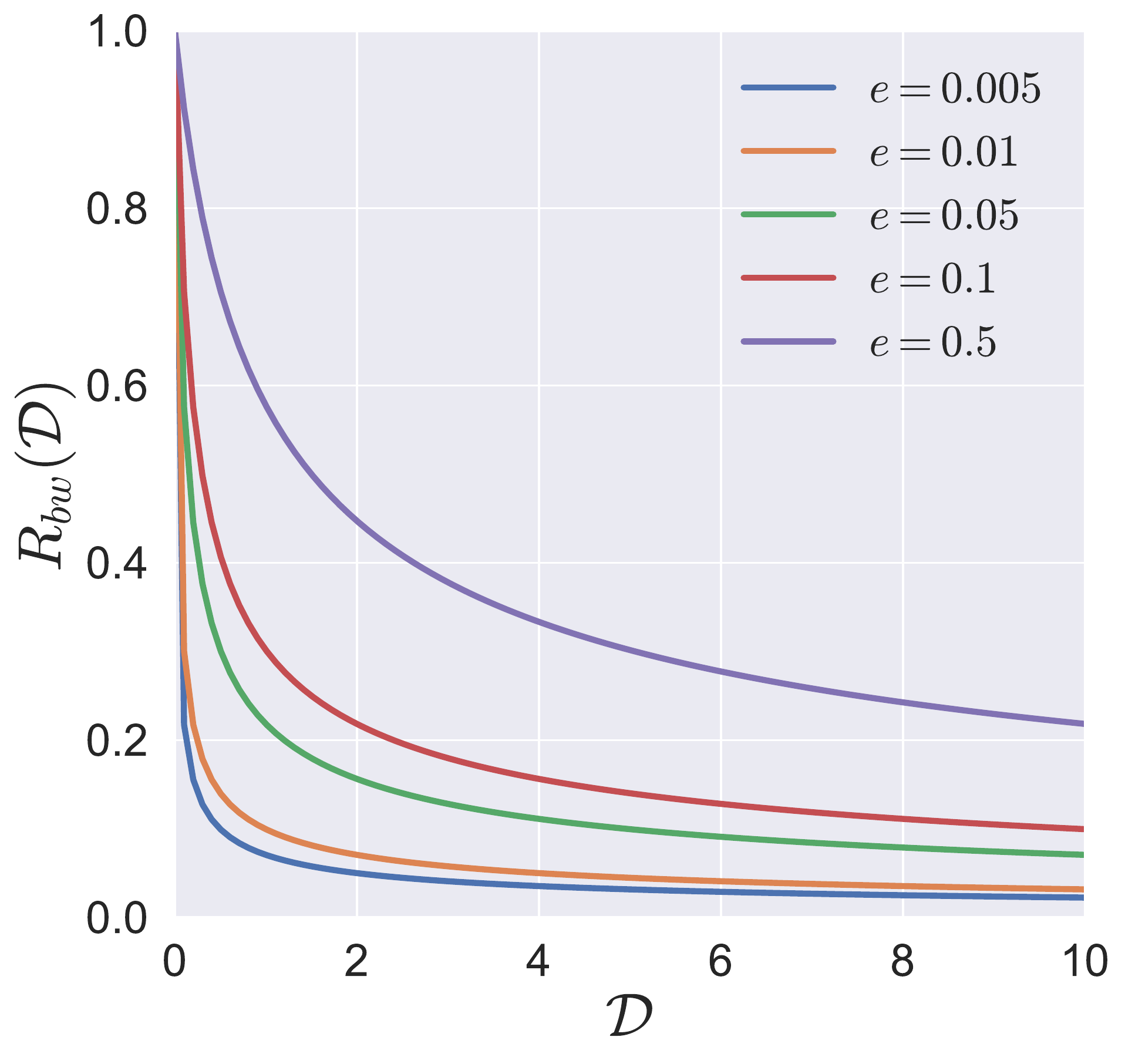}\label{fig:Rbw_2dof_coupled_depth}}
	\subfigure[]{\includegraphics[width=0.48\textwidth]{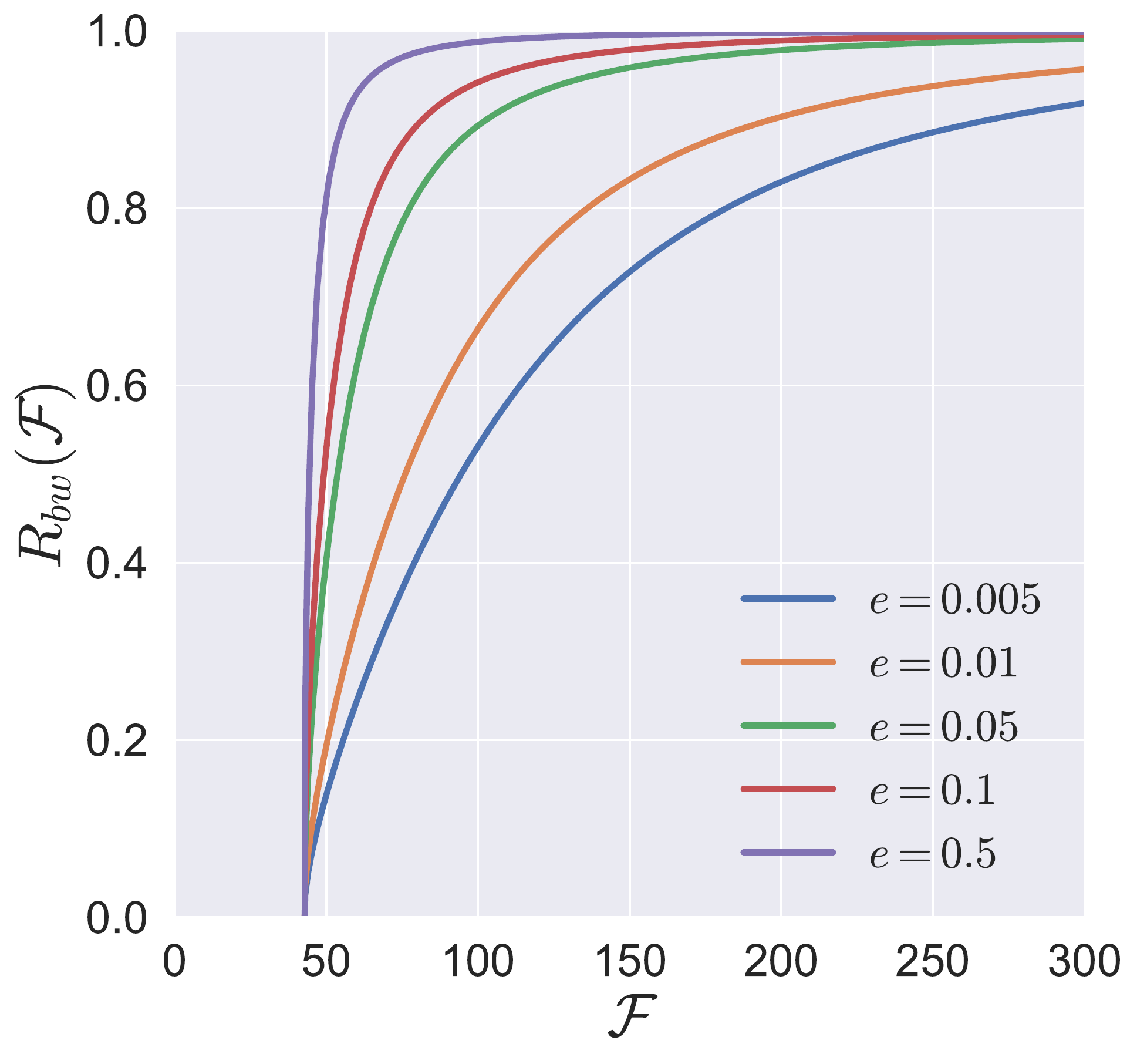}\label{fig:Rbw_2dof_coupled_flatness}}
	\caption{\textbf{Ratio of the bottleneck-width and well-width for the two DOF coupled system.} Shows the changes in the ratio for increasing (a) depth and (b) flatness of the PES at different total energies. System parameters $\mu = 4,\omega = 3,\varepsilon = 5$ are fixed, while the depth and flatness of the PES is changed by varying the parameter $\alpha$.}
	\label{fig:Rbw_2dof_coupled}
\end{figure}

For the two DOF coupled system, that is $\varepsilon \neq 0$, we use the same procedure as described above. 
Thus, the ratio of the bottleneck-width and well-width, $R_{bw}$, becomes 
\begin{align}
R_{bw} = \dfrac{\sqrt{\dfrac{2e}{\omega^2+ \varepsilon}}}{\sqrt{A|_{x=x_{\rm cent},V(x,y)=e}} \sqrt{\dfrac{2}{\omega^2 + \varepsilon}}} = \sqrt{\dfrac{e}{e + \mathcal{D_{\varepsilon}}}}
\end{align}
where $A$ is defined in the appendix and where $\mathcal{D_{\varepsilon}}$ is defined in Eqn.~\eqref{eqn:depth_sn2dof_coupled}. This relationship of the ratio, $R_{bw}$, and depth is shown in the Fig.~\ref{fig:Rbw_2dof_coupled_depth}. To show the influence of the flatness on this ratio, we evaluate the expression~\eqref{eqn:flatness_sn2dof} numerically for different values of the system parameters and show the changes in terms of the flatness measure in Fig.~\ref{fig:Rbw_2dof_coupled_flatness}. However, the influence of the flatness on this ratio for small values is entirely absent until the flatness increases above $\approx 50$ after which we observe a nonlinear monotonic growth that asymptotes towards equal bottleneck-width and well-width.

\begin{figure*}[!ht]
	\centering
	\subfigure[]{\includegraphics[width=0.33\textwidth]{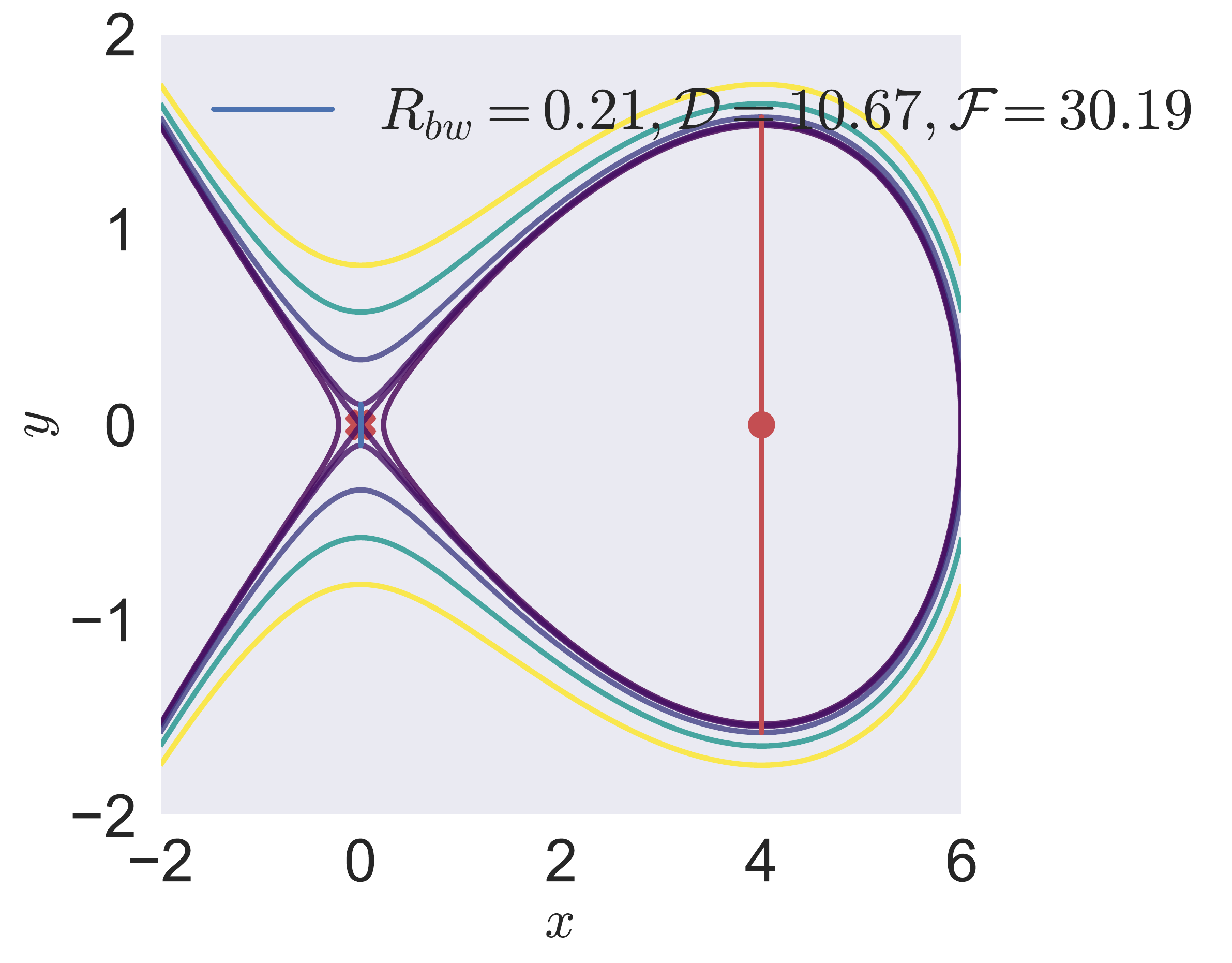}}
	\subfigure[]{\includegraphics[width=0.33\textwidth]{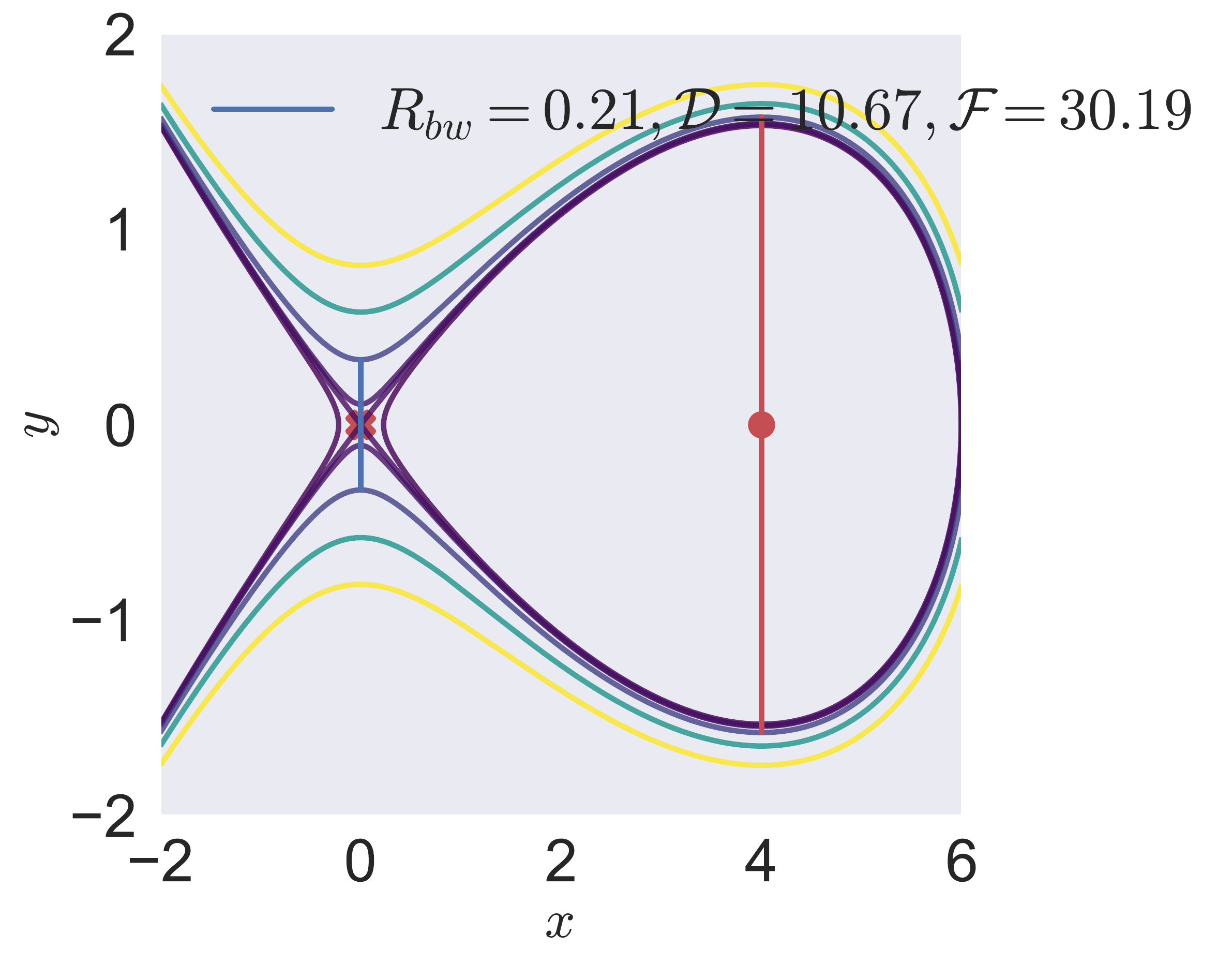}}
	\subfigure[]{\includegraphics[width=0.33\textwidth]{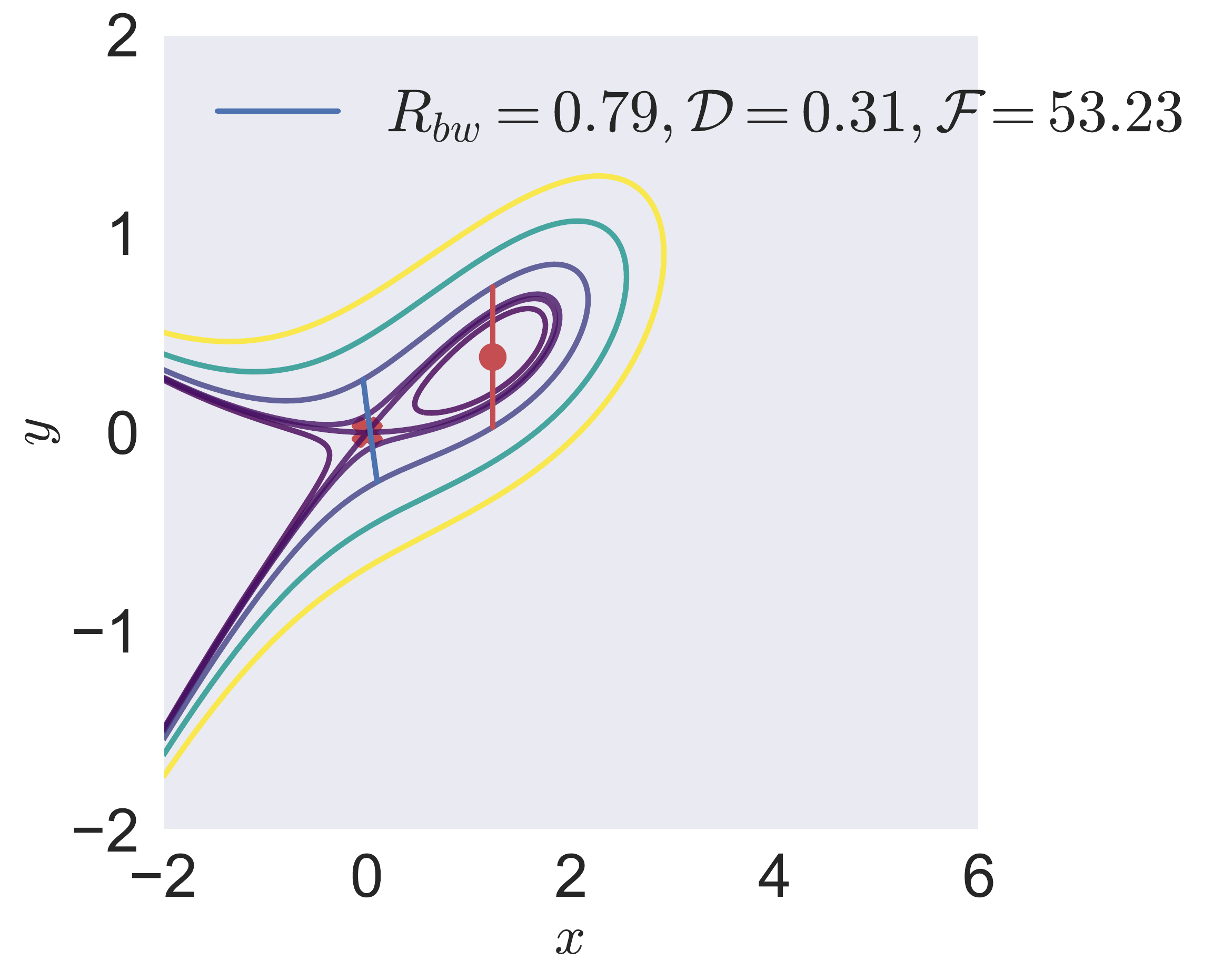}}
	\caption{\textbf{Configuration space view of the ratio of the bottleneck-width and well-width.} Shows the equipotential contours for the two DOF system. (a) $\varepsilon=0, \ \Delta E = 0.05$ and $R_{bw} = 0.07$ (b) $\varepsilon=0, \ \Delta E = 0.5$, and $R_{bw } = 0.21$  (c) $ \varepsilon = 4.0, \ \Delta E=0.5$, and $R_{bw } = 0.79$. In all the plots, parameters $\alpha=1.0, \ \mu=4.0, \ \omega=3.0$ are fixed.}\label{fig:Rbw_viz_sn2dof}
\end{figure*}

Thus, we see that the ratio of the bottleneck-width and well-width can be summarized by $R_{bw}=\sqrt{e/(e + \mathcal{D})}$ which depends on the total energy of the system and the depth of the PES in the saddle-node Hamiltonian. The qualitative changes in the bottleneck-width and the well-width in the configuration space of the two DOF systems can be visualized in Fig.~\ref{fig:Rbw_viz_sn2dof}.

We note that the multivalued ratio for small flatness in the one and two DOF uncoupled systems is because the depth and small flatness does not have a one-to-one mapping for the 1 DOF and 2 DOF system with small coupling. We can see from Fig 2 (a),(b) that the same value of flatness corresponds to two different depth values, and thus two different values of the ratio, $R_{bw}$, in the flatness plots.



\subsection{Reaction probability}

We define the reaction probability at time $t$ as the fraction of \emph{reactive} trajectories at a given total energy, $e$. To calculate this measure of reaction, we sample points in the reactants region with the fixed energy constraint, $\mathcal{H}(x,y,p_x,p_y) = e$. We then let those points evolve in the phase space and count how many of them have reached the products region. Trajectories that reach the products region at time $t$ are reactive and the rest in the sample are nonreactive trajectories. 

\subsubsection{One degree-of-freedom Hamiltonian}

The phase space region $x > 0$ is defined as the reactants region and $x < 0$ as the products region. Given an initial condition in $x > 0$ region, reaction occurs when the trajectory goes through the forward dividing surface (DS) given by the Eqn.~\ref{eqn:forw_DS_sn1dof} and initial condition in $x < 0$ region will enter the reactant by passing through the backward DS (Eqn.~\ref{eqn:back_DS_sn1dof}). For this system, all initial conditions above the energy of the saddle will react eventually by passing through the DS given by:
\begin{align}
\text{ forward DS:} \quad \, \left\{ (x,p_x) \in \mathbb{R}^2 \, | \, x = 0, p_x = - \sqrt{ 2e} \right\} \ \label{eqn:forw_DS_sn1dof} \\
\text{ backward DS:} \quad \, \left\{ (x,p_x) \in \mathbb{R}^2 \, | \, x = 0, p_x = + \sqrt{ 2e} \right\} \  \label{eqn:back_DS_sn1dof}
\end{align}

\begin{figure}[!ht]
	\centering
	\subfigure[]{\includegraphics[width=0.49\linewidth]{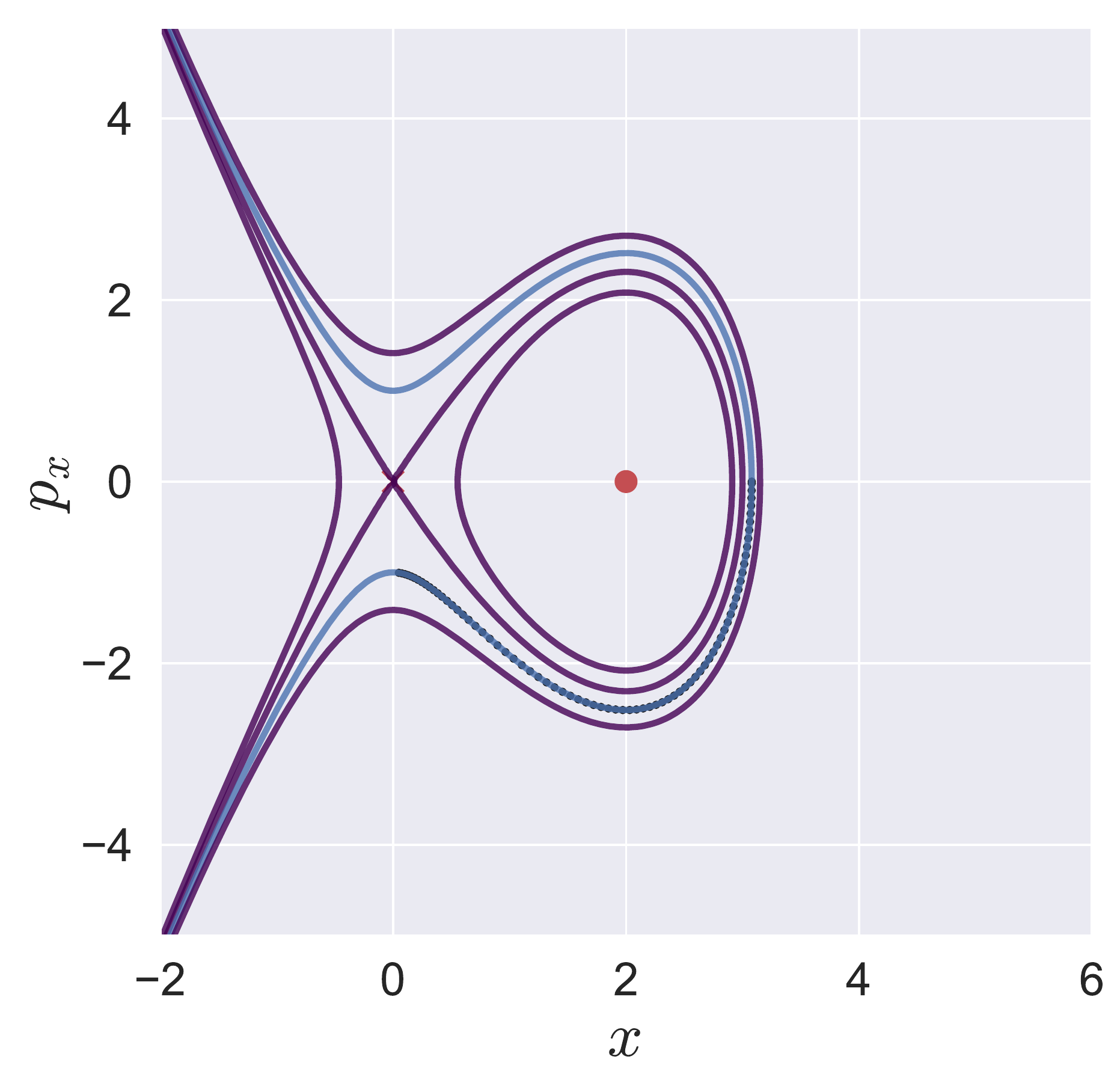}\label{fig:sample_pos_sn1dof}}
	\subfigure[]{\includegraphics[width=0.49\linewidth]{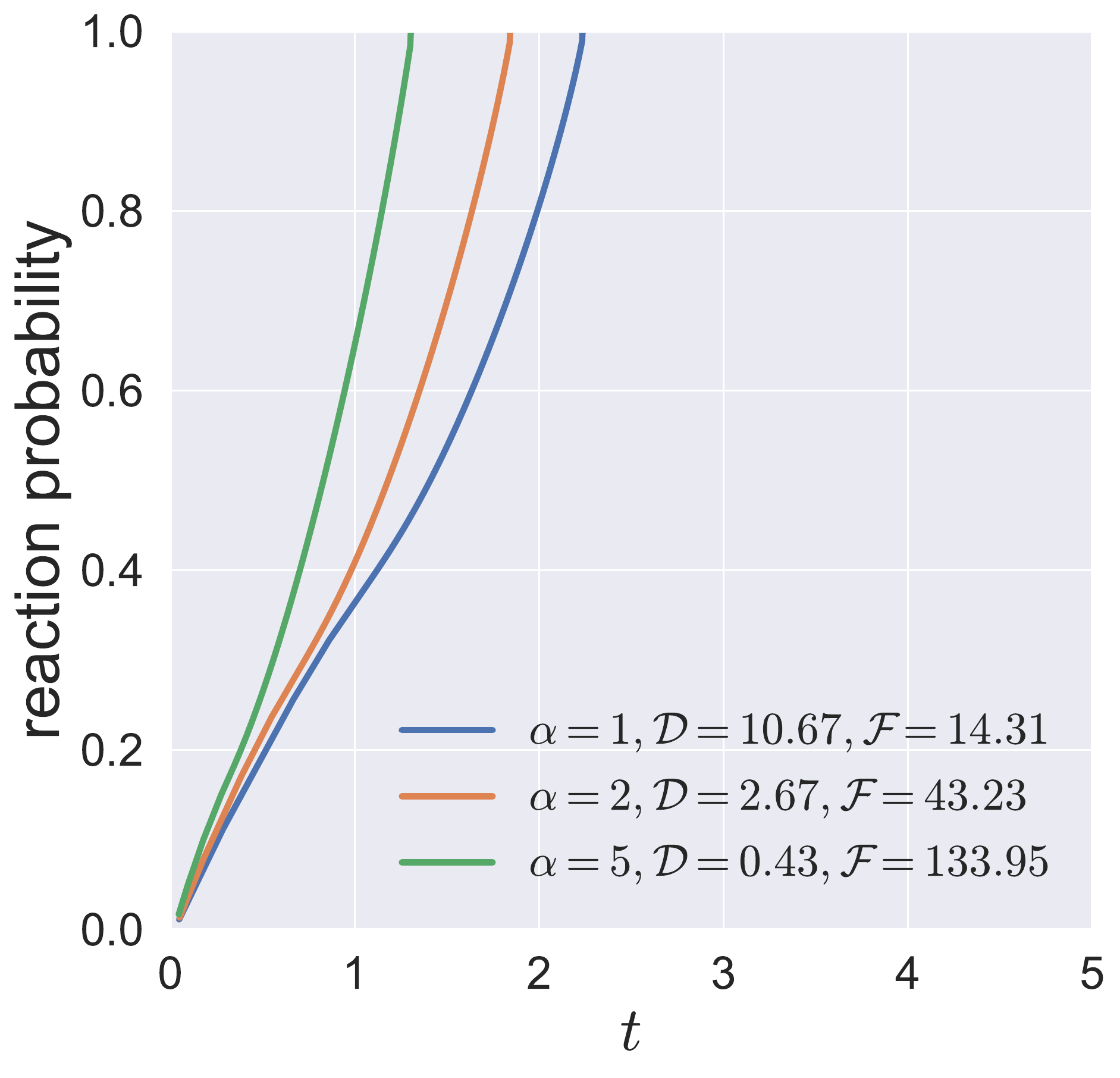}\label{fig:react_prob_sn1dof}}
	\caption{\textbf{Reaction probability for the one DOF Hamiltonian}. (a) The red ``cross'' marks the location of the saddle point, the red ``dot'' marks the location of the center point. The black circles show the microcanonical ensemble on the isoenergetic contour at total energy, $e = 0.5$ in blue. $\alpha = 2$ is fixed. (b) Shows the increase in the reaction probability with decreasing depth and increasing flatness, amd $\mu = 4$ is fixed.}
	\label{fig:samp_pos_reactprob_1dof}
\end{figure}

To estimate the reaction probability, we perform a Monte Carlo simulation of a microcanonical ensemble of initial conditions in the two dimensional phase space define by the constraint 
\begin{equation}
    \mathcal{L}^- = \left\{ (x,p_x) \in \mathbb{R}^2 \, | \, x > 0, p_x(x;e) < 0 \right\}
\end{equation}
These initial conditions are shown as circles on the isoenergetic contour corresponding to the total energy, $e = 0.5$ in the Fig.~\ref{fig:sample_pos_sn1dof}.

The results of the Monte Carlo simulation is shown in the Fig.~\ref{fig:react_prob_sn1dof}, we can see that all trajectories react (since the reaction probability equals unity) within 3 time units across all the depth and flatness values considered. We can see that the curve for $\alpha = 1,2$ has two different slopes (the curve seems linear until $t = 1$ and then becomes nonlinear) and this biphasic slope vanishes for $\alpha = 5$. We can also see that if we decrease the depth of the PES and increase the flatness of the PES, that is the PES becomes less deep and more flat, reaction probability increases and at a faster rate.


\subsubsection{Two degree-of-freedom Hamiltonian}

In the two DOF system, we define the phase space region $x>0$ as the reactants region and $x<0$ as the products region. Given an initial condition with $x > 0$, the reaction occurs when the trajectory goes through the periodic orbit (unstable) dividing surface (for 2 DOF systems, it is a 2-sphere, $\mathbb{S}^2$) constructed in the phase space~\cite{waalkens2004direct}. We note here that, unlike the one DOF system, not all trajectories will lead to reaction and there will be trajectories that remain trapped in the reactants region. This observation can be linked to the dynamical trapping due to heteroclinic intersections of the invariant manifolds~\cite{katsanikas_dynamical_2020} or due to the presence of KAM tori (phase space regions of regular motion). 

For the uncoupled system, the dividing surface constructed from the unstable periodic orbit (which is the NHIM for a 2 DOF system) is defined by the condition $x = 0$ and becomes
\begin{align}
\text{ forward DS:} \quad & \left\{ (x,y,p_x,p_y) \in \mathbb{R}^4 \, | \, x = 0, p_x(y,p_y;e) < 0 \right\} \label{eqn:forwds_sn2dof_uncoupled}\\
\text{ backward DS:} \quad & \left\{ (x,y,p_x,p_y) \in \mathbb{R}^4 \, | \, x = 0, p_x(y,p_y;e) > 0 \right\} \label{eqn:backds_sn2dof_uncoupled}
\end{align}
where $e$ is the total energy of the system and $p_x(y,p_y;e)  = \pm \sqrt{ 2e - p_y^2  -\omega^2 y^2}$ is the fixed energy constraint. The forward reaction occurs when the trajectory crosses the forward DS and the backward reaction occurs when the trajectory crosses the backward DS. We note here that this dividing surface constructed in the phase space has the locally no-recrossing property~\cite{waalkens2004direct}, and in general, trajectories will show global recrossings of the DS due to the Poincar{\'e} recurrence theorem~\cite{wiggins2003applied}. However, since the energy surface in the two DOF saddle-node Hamiltonian is unbounded (open potential well), trajectories that go beyond a certain negative $x-$coordinate do not return to cross the DS. 

\begin{figure}[!ht]
	\centering
	\subfigure[]{\includegraphics[width=0.23\textwidth]{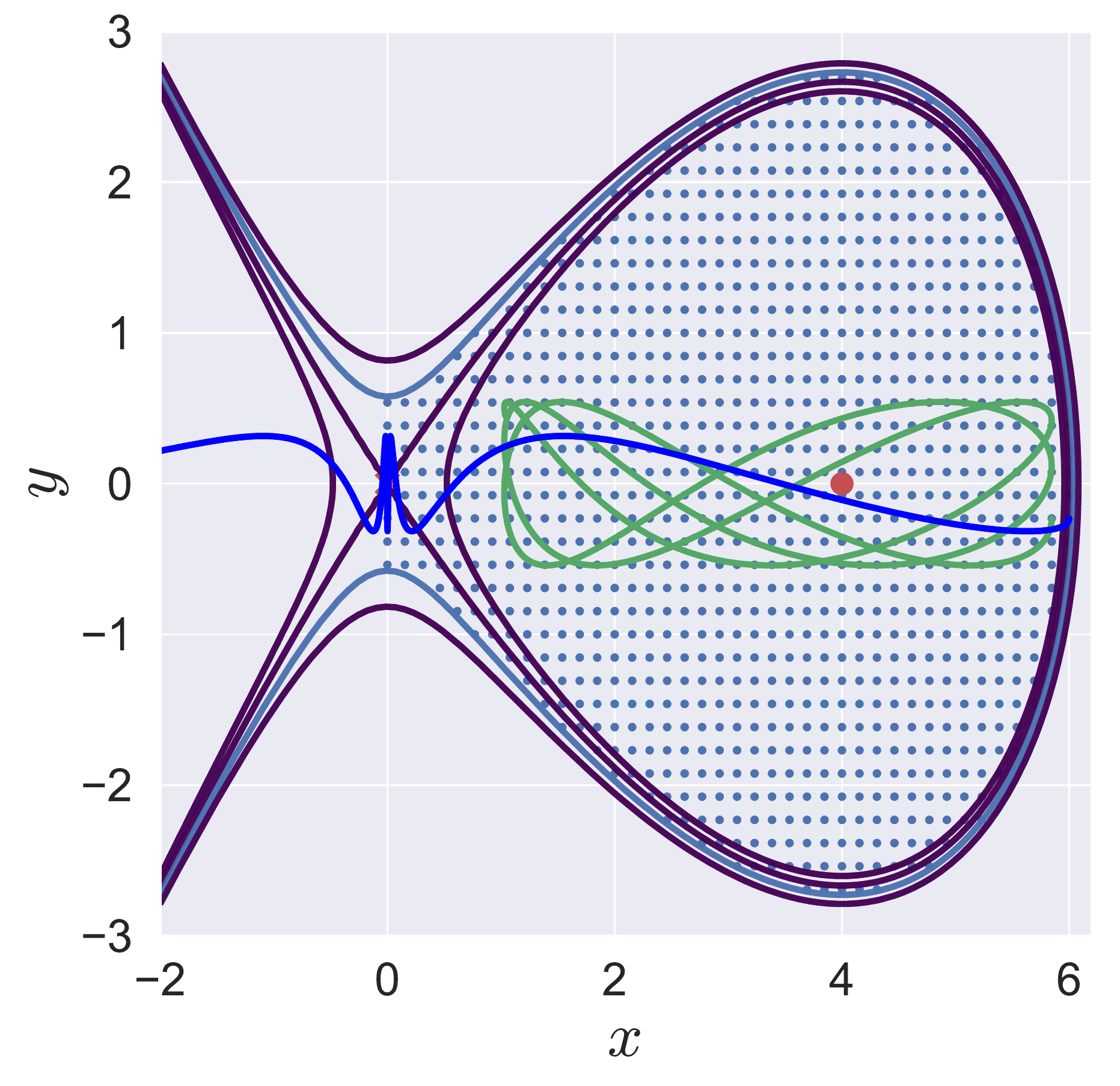}}
	\subfigure[]{\includegraphics[width=0.23\textwidth]{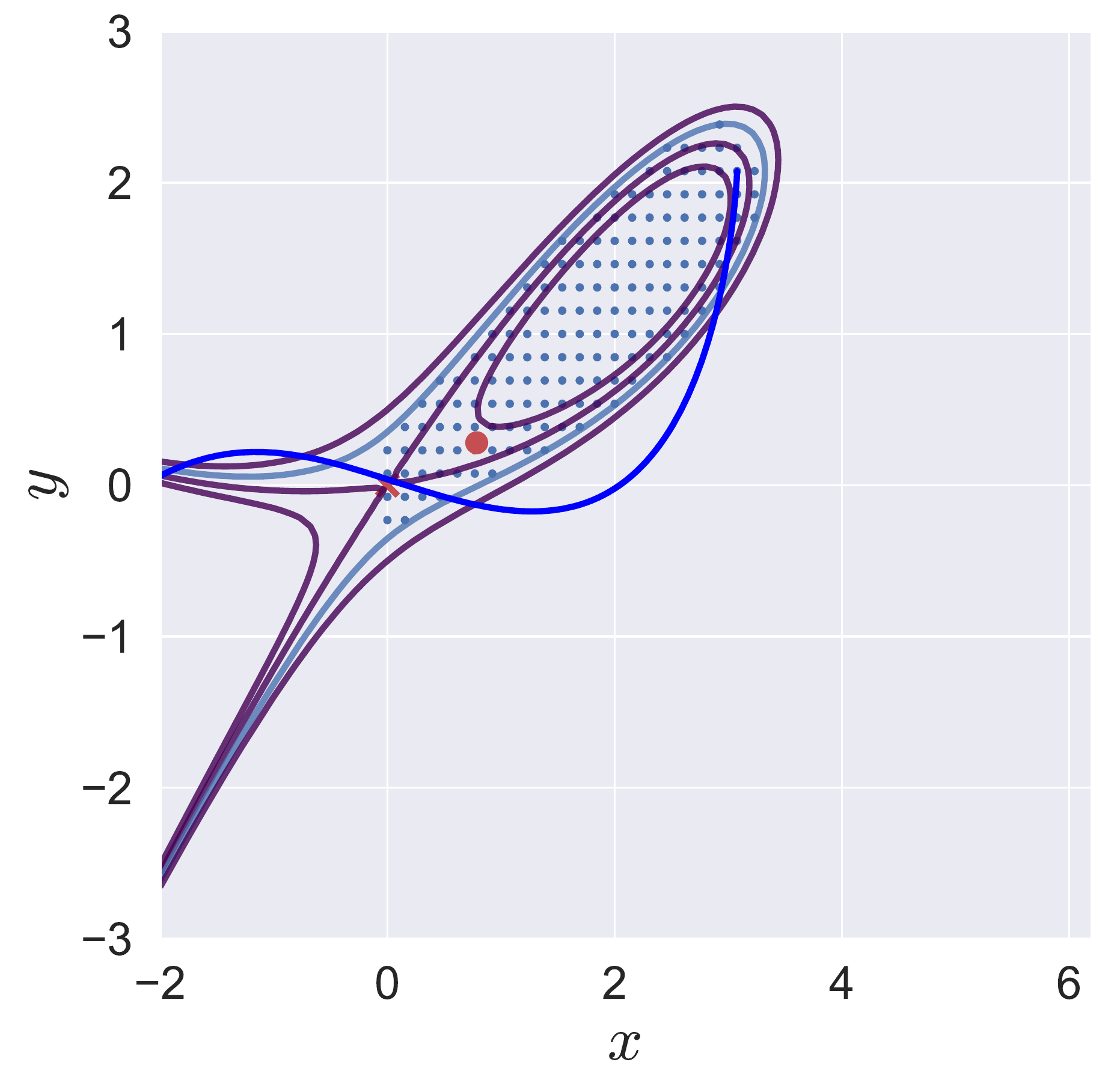}}
	\caption{\textbf{Configuration space view of the microcanonical ensemble and equipotential contours} in the two DOF saddle-node Hamiltonian with parameters: $\alpha=1,\mu=4,\omega=3, e=0.5$, (a) uncoupled: $\varepsilon=0$, and (b) coupled: $5$. Since the reaction is defined as going from $x > 0$ to $x < 0$, trajectories are initialized with $p_x < 0$. The blue vertical line crosses the origin is the DS, the blue curve is the projection of the PES with total energy 0.5 in the configuration space. The red ``cross'' sign is the location of the saddle point, the red ``dot'' sign is the location of the centre point. The blue points correspond to the choose of initial conditions. The blue curves are reactive trajectories and the green curve is a nonreactive trajectory.}\label{fig:samp_pos_reactprob_2dof}
\end{figure}

For the coupled system, we do not have an explicit analytical form for the periodic orbit based DS since the unstable periodic orbit has to be computed using a turning point or continuation type method~\cite{Lyu2020}. However, the DS is given by
\begin{align}
\text{ forward DS:} & \; \left\{ (x,y,p_x,p_y) \in \mathbb{R}^4 \, \vert \, p_x(x_{\rm UPO},y_{\rm UPO},p_y;e) < 0 \right\} \label{eqn:forwds_sn2dof_coupled}\\
\text{ backward DS:} & \; \left\{ (x,y,p_x,p_y) \in \mathbb{R}^4 \, | \, p_x(x_{\rm UPO},y_{\rm UPO},p_y;e) > 0 \right\}\label{eqn:backds_sn2dof_coupled}
\end{align}
where $p_x  = \pm \sqrt{ 2e - p_y^2  -\omega^2 y^2 + \sqrt{\mu} x^2 - (\alpha/3) x^3}$ is the fixed energy constraint and $x_{\rm UPO}, y_{\rm UPO} \in \mathrm{UPO}$. To simplify our computation, we sample points in the phase space region with $x > 0$ and use the line $x = -5$ as the fictitious boundary of the phase space, that is trajectories are stopped once they cross the line $x = -5$.  This condition ensures that all reactive trajectories pass through the recrossing free DS (Eqns.~\eqref{eqn:forwds_sn2dof_coupled} and~\eqref{eqn:backds_sn2dof_coupled}) and are terminated thereafter. A representative reactive trajectory and the distribution of sample initial conditions on the energy surface is shown as the projection on the configuration space in Fig.~\ref{fig:samp_pos_reactprob_2dof}. 
In Fig.~\ref{fig:samp_pos_reactprob_2dof}, the blue curves are two representative reactive trajectories which start from one of the sampling points with negative $p_x$ and total energy $e=0.5$. We can see that they start from the reactants region, cross through the DS and escape to the products region. The green curve is a representative nonreactive trajectory which starts from one of the sampling points with negative $p_x$ and total energy $0.5$. We can see that the trajectory is trapped in the reactants region and will not evolve to the products region.

\begin{figure}[!ht]
	\centering
	\subfigure[\label{fig:reactprob_2dof_ep0}]{\includegraphics[width=0.45\textwidth]{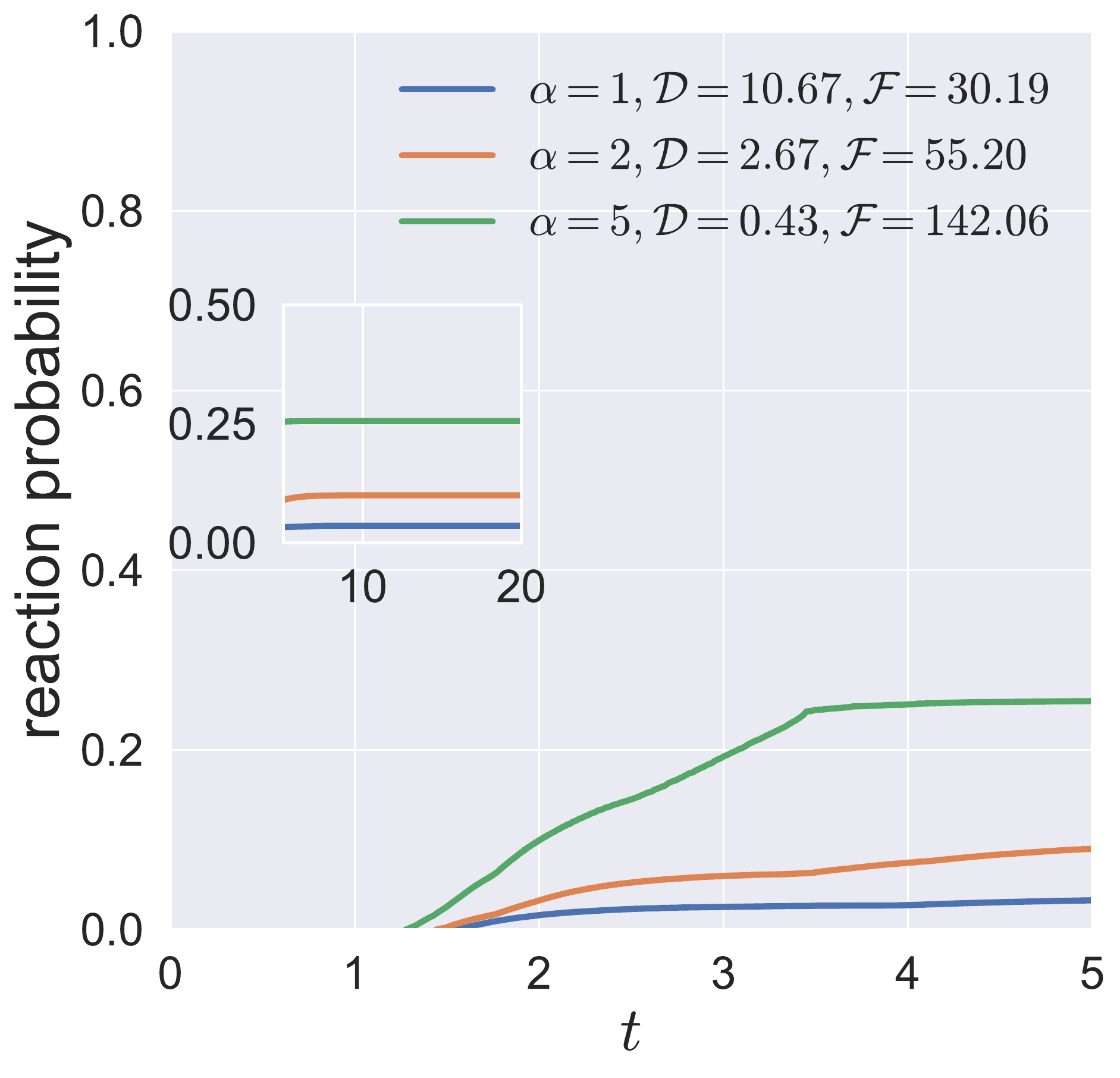}}
	\subfigure[\label{fig:reactprob_2dof_ep5}]{\includegraphics[width=0.45\textwidth]{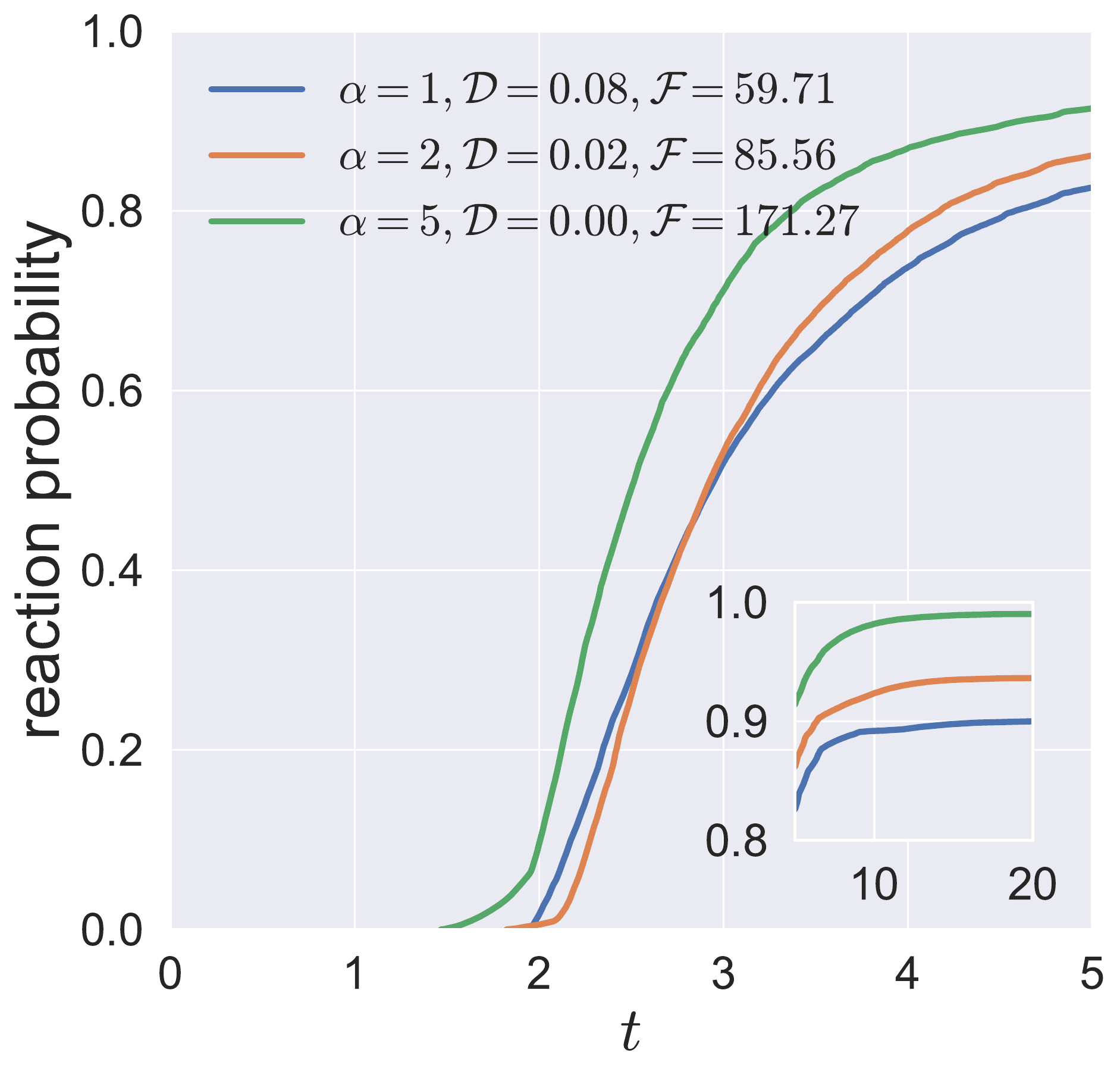}}
	\caption{\textbf{Reaction Probability for the two DOF system.} (a) uncoupled: $\varepsilon=0$ and (b) coupled: $\varepsilon=5$.  In both systems, the reaction probability increases with decreasing depth and increasing flatness. While the coupled system shows almost all trajectories lead to reaction, the uncoupled system shows trajectories stay trapped in the well even though their initial momentum, $p_x$, was directed towards the bottleneck. $\mu = 4, \ \omega=3, \ e=0.5$ are fixed for both systems.}
	\label{fig:reactprob_2dof}
\end{figure}

We perform a Monte Carlo simulation of the microcanonical ensemble of initial conditions in the four dimensional phase space with $p_x < 0$. First, we select approximately 100 initial points $\left\{x_i,y_i\right\}$ in the configuration space, and these points satisfy the condition $V(x_i,y_i) \leq e,\ x_i \geq 0$ and where $e$ is the total energy of system. For each point $\left\{x_i,y_i\right\}$, we then select 100 random values of negative $p_x$. This ensemble is integrated until they cross the fictitious boundary for both the two DOF uncoupled and coupled systems.The reaction probability obtained by varying the depth, flatness, and total energy are shown in Fig.~\ref{fig:reactprob_2dof}.
Comparing Fig.~\ref{fig:reactprob_2dof} (a) and (b), we see the effect of coupling on the reaction probability magnitudes and rate of growth. For the uncoupled system in Fig.~\ref{fig:reactprob_2dof} (a), the reaction probabilities increase gradually (compared to the coupled system in Fig.~\ref{fig:reactprob_2dof} (b)) during the time interval $[0,10]$. After time $t=10$, the reaction probabilities remain constant and only $25\%$ or less (depending on the depth and flatness) trajectories are reactive trajectories. A different behavior is observed for the coupled system in Fig.~\ref{fig:reactprob_2dof} (b) and more than $80\%$ (depending on the depth and flatness) of trajectories are reactive trajectories. Across all the plots in Fig.~\ref{fig:reactprob_2dof}, we observe that the reaction probability increases with decreasing depth and increasing flatness, except in the coupled case where there is a crossing of the reaction probabilities for small depths and high flatness over the interval, $2.5 < t < 3.0$. Aside from this peculiar crossing, these calculations show that if we decrease the depth of the PES and increase the flatness of the PES, that is the PES becomes less deep and more flat, more trajectories lead to reaction for reach the products region and react.




\subsection{Gap time distribution}

We adopt the definition of the gap time~\cite{ezra_microcanonical_2009} which is the time between the two successive recrossings of the DS. To estimate this quantity, we start on the DS with $p_x > 0$ such that the trajectory enters the reactants region, then the time instant when it recrosses the DS is the gap time or the first passage time. For our system, when a trajecotory recrosses the DS and enters the products region, the sign of $p_x$ changes to negative and $x < 0$. This happens when the trajectory crosses the forward DS given by the Eqn.~\eqref{eqn:forwds_sn2dof_coupled}. Due to the unbounded energy surface, trajectories that go beyond the fictitious boundary do not return to the neighborhood of the DS. We perform this calculation for a microcanonical ensemble of initial conditions and we record the gap times for all the trajectories starting on the DS with positive $p_x$, and call this the microcanonical gap time distribution~\cite{ezra_microcanonical_2009}. In this subsection, we discuss our results on the gap time distribution by varying the depth and flatness using the parameter, $\alpha$, at a fixed total energy.


\subsubsection{One degree-of-freedom Hamiltonian}

For the one DOF system, the DS consists of two points defined in the Eqn.~\ref{eqn:forw_DS_sn1dof} and~\ref{eqn:back_DS_sn1dof}. The gap time is the time when trajectory starts on the initial position $x=0,p_x = + \sqrt{ 2e}$ and reaches the final position $x=0,p_x = - \sqrt{ 2e}$. Therefore, the gap time is a single value and is given by the time taken to move along the isoenergetic contour at $\mathcal{H}(x,p_x) = e$.

\begin{figure}[!ht]
	\centering
	\subfigure[]{\includegraphics[width=0.23\textwidth]{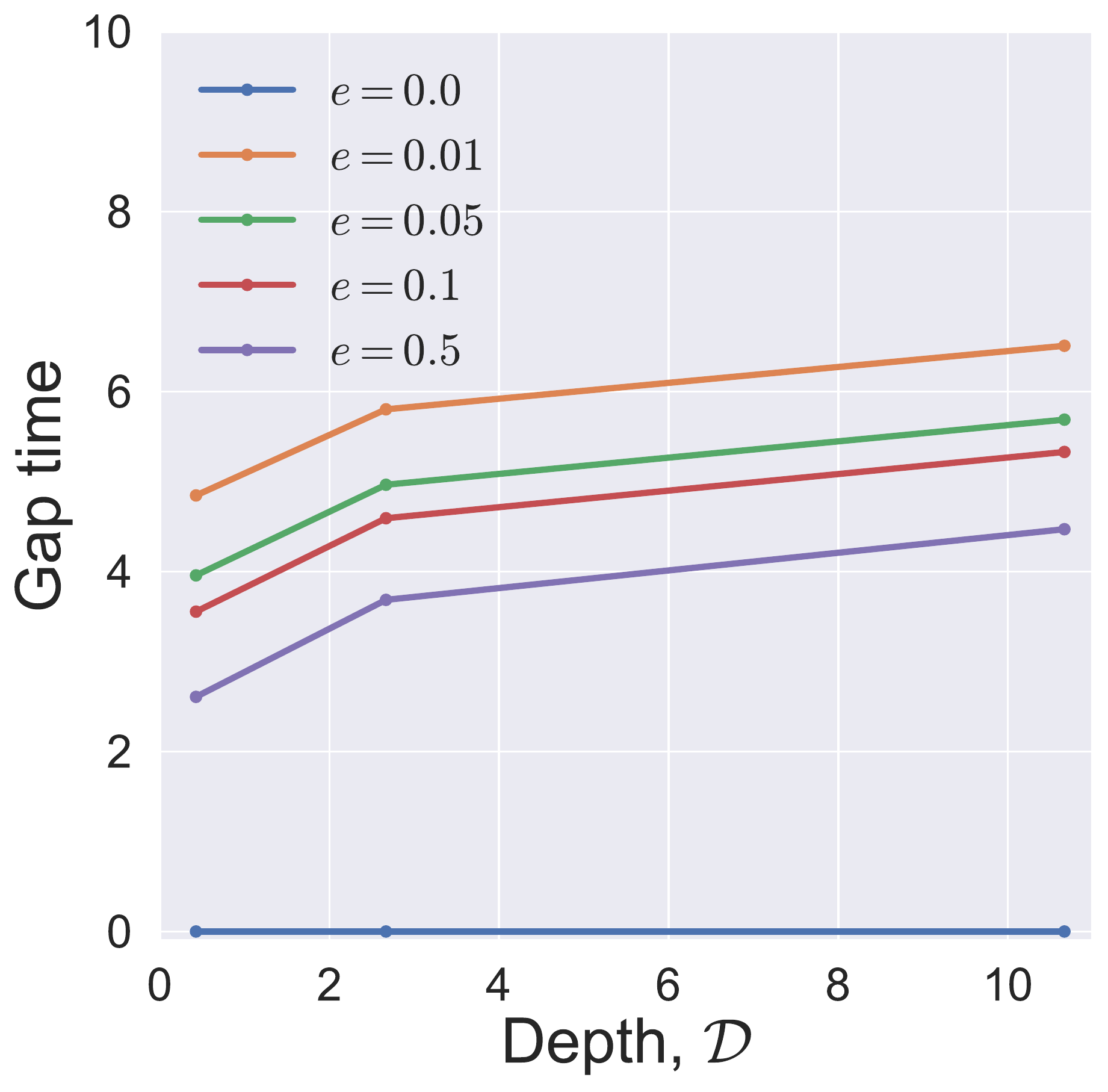}\label{fig:gaptime_dep_1dof}}
	\subfigure[]{\includegraphics[width=0.23\textwidth]{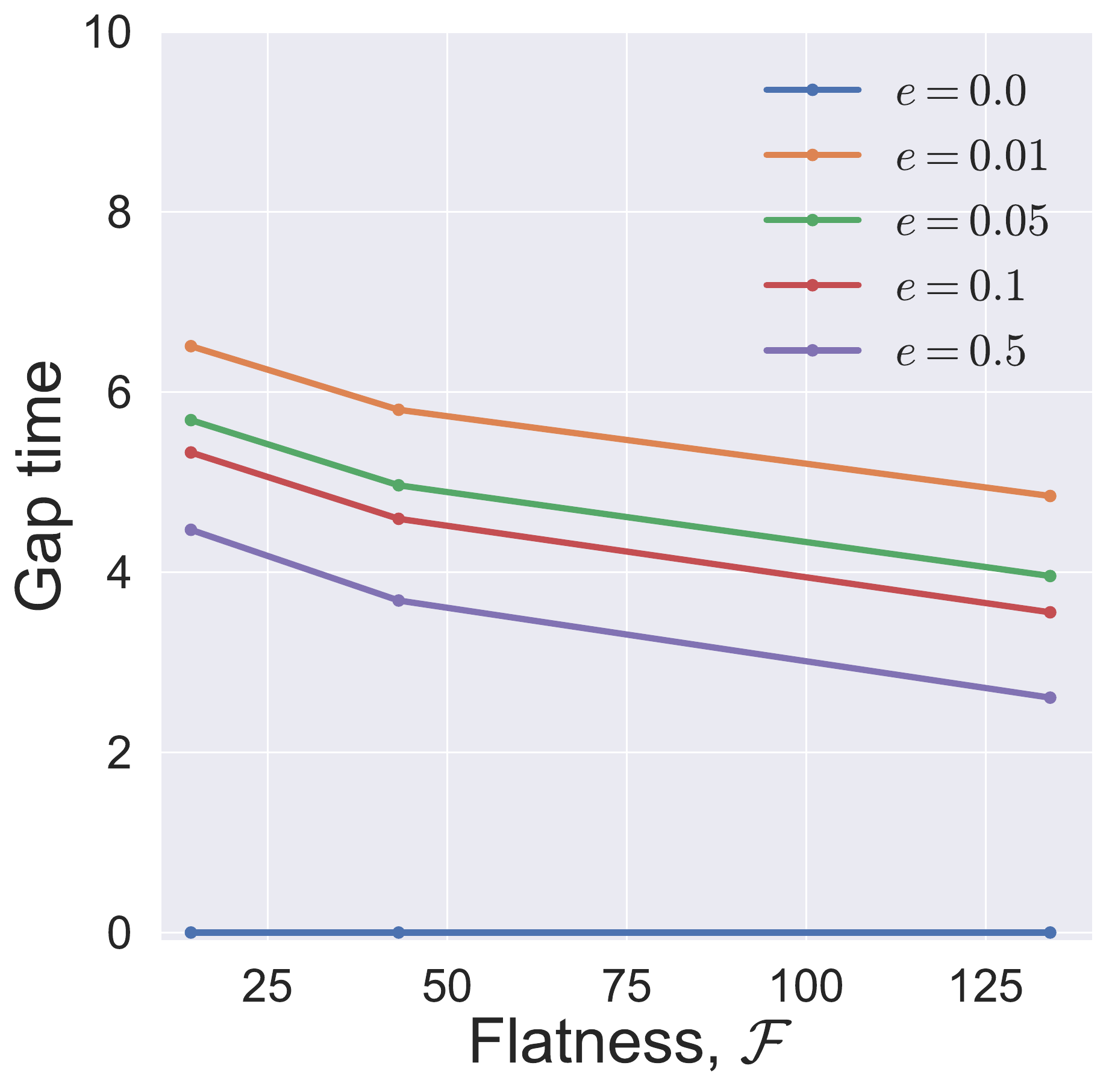}\label{fig:gaptime_flat_1dof}}
	\caption{\textbf{Gap times for the one DOF system.} The depth and flatness for each line is varied using $\alpha$ and correspond to the values $\alpha =1: (\mathcal{D},\mathcal{F})=(10.67,30.19)$, $\alpha =2: (\mathcal{D},\mathcal{F})=(2.67,43.23)$, and $\alpha =5: (\mathcal{D},\mathcal{F})=(0.43,133.95)$. $\mu = 4$ is fixed in all the cases.}
	\label{fig:gaptime_1dof}
\end{figure}


\subsubsection{Two degree-of-freedom Hamiltonian} 


In this subsection, we use the algorithm~\cite{Ezra_sampling2018} to sample points on the phase space DS (which is a 2-sphere) for calculating the gap time distribution. First, we compute the NHIM which is an unstable periodic orbit associated with the index-1 saddle equilibrium point in a two DOF system at a total energy $e$.  Second, we collect the configuration space coordinates $\{(x_i,y_i)\}$ on the NHIM. This corresponds to projecting the unstable periodic orbit (UPO at energy $e$) onto the configuration space. For each $\{(x_i,y_i)\}$, our system in the $(p_x, p_y)$ plane is given by:
\begin{equation}
\frac{1}{2}(p_x^2+p_y^2) = e - V(x_i,y_i)
\end{equation}
Therefor, the maximum value of $p_x$ is
\begin{equation}
p_x^{max} = \sqrt{2(e - V(x_i,y_i))}
\end{equation}
and we select $p_{x,i}$ uniformly from the interval $[-p_x^{max},p_x^{max}]$. Using the definition of the Hamiltonian, we calculate the value of $p_{y,i}$, which is either positive or negative. We note that $p_y^{max} = p_x^{max},\ p_y \in \left[-p_y^{max},p_y^{max}\right]$.

\begin{figure}[!ht]
	\centering
	\subfigure[]{\includegraphics[width=0.45\textwidth]{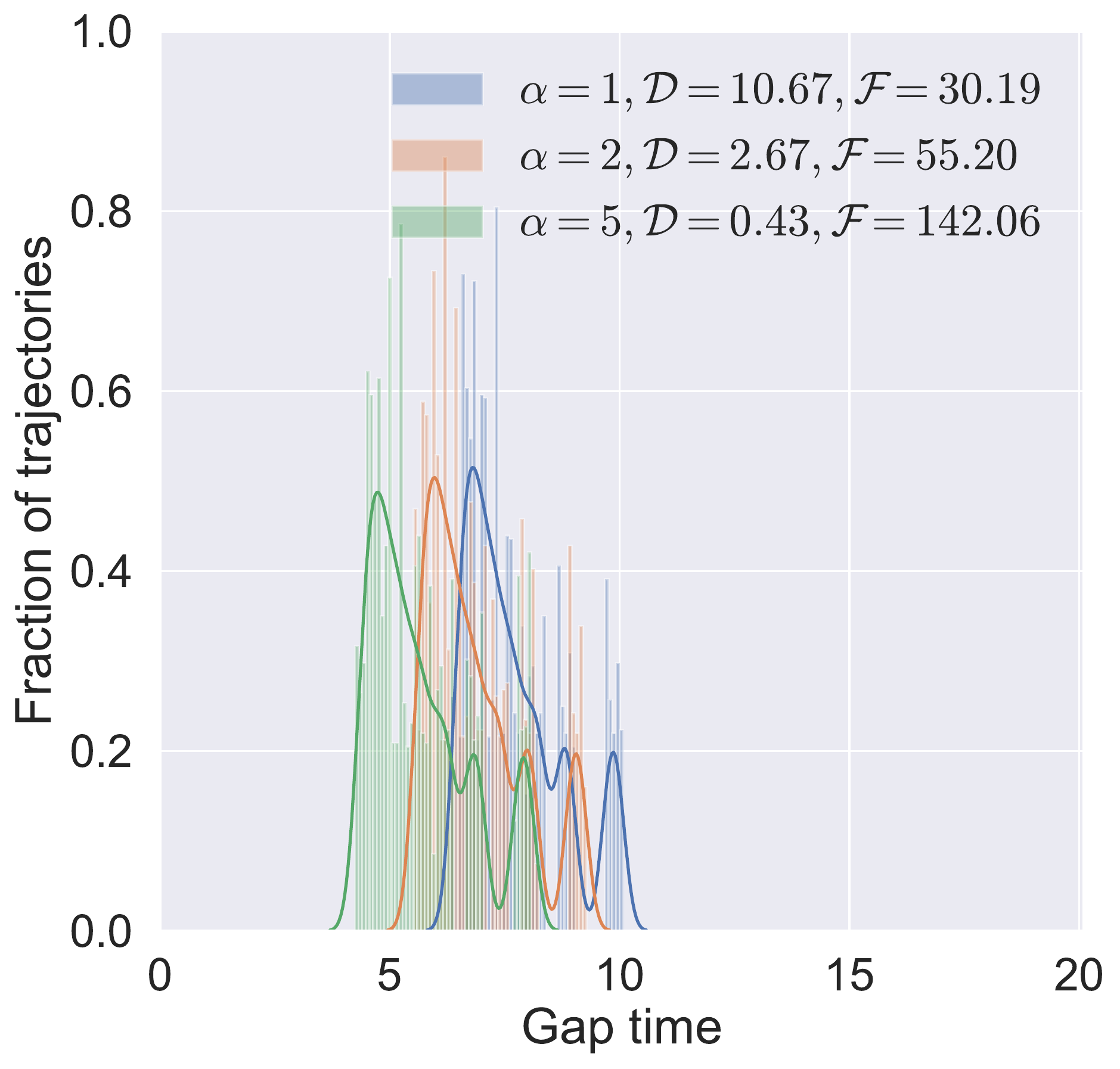}\label{fig:gaptime_uncoupled_2dof}}
	\subfigure[]{\includegraphics[width=0.45\textwidth]{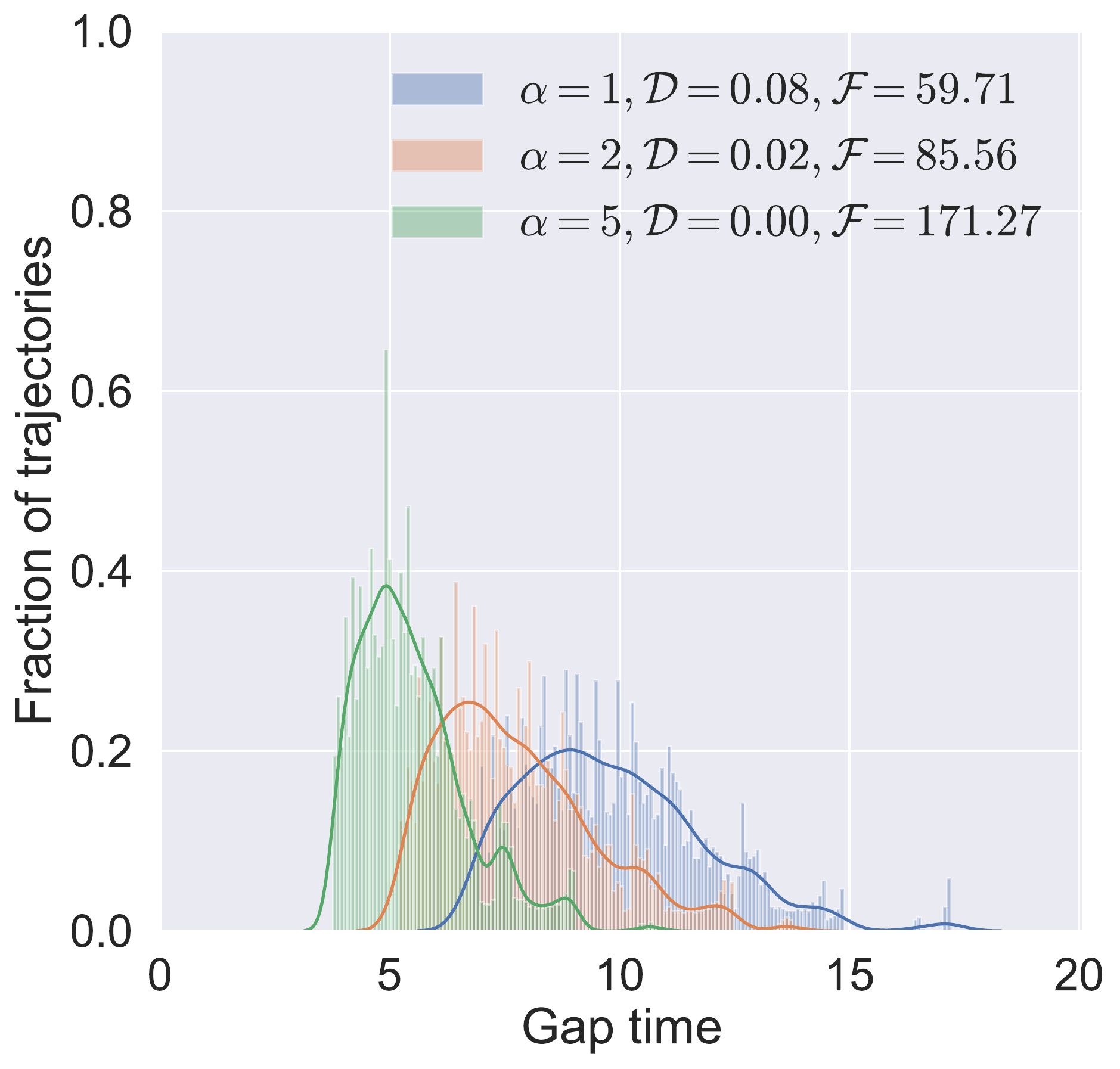}\label{fig:gaptime_coupled_2dof}}
	\caption{\textbf{Gap times for the two DOF system.} Shows the distribution for the (a) uncoupled: $\varepsilon = 0$ and (b) coupled: $\varepsilon = 5$ system. In these plots, the depth and flatness is varied using $\alpha$ and shows the temporal shift (decreasing) in the mode of the distribution with decreasing depth and increasing flatness. $\mu = 4,\omega = 3, e = 0.5$ are fixed in all the cases.}
	\label{fig:gaptime_2dof}
\end{figure}

Using this algorithm, we generate a set of microcanonical ensemble $\{( x_i,y_i,p_{x,i},p_{y,i})\}_e$ on the phase space DS at a total energy $e$. We select the initial conditions with positive $p_x$ and integrate so that trajectories enter the reactants region, spend time in the reactants region and finally leave the reactants region. We record the time when a trajectory leaves the reactants region as the gap time of the initial condition of the trajectory. For the coupled system, the UPO needs to be computed using numerical method~\cite{Lyu2020} and it has been shown~\cite{garciagarrido_tilting_2019} that for the projection onto the configuration space tilts with the changes in the parameters of the PES. Thus, to simplify this detection of crossing the forward DS, we use the fictitious boundary condition $x = -5$ which ensures that the trajectory has left the reactants region and is in the products region. The gap time distributions are shown in Fig.~\ref{fig:gaptime_uncoupled_2dof} and~\ref{fig:gaptime_coupled_2dof} for the uncoupled and coupled systems, respectively.

For the uncoupled system, in Fig.~\ref{fig:gaptime_2dof}\protect\subref{fig:gaptime_uncoupled_2dof}, we see that the gap time distributions have similar shapes for the three values of the depth and flatness that we considered; the distribution only shifts to a later time as the depth is increased and flatness is decreased. Thus, if we decrease the depth of the PES and increase the flatness of the PES, that is the PES becomes less deep and more flat, trajectories spend less time in the reactants region; an established observation for chemical reactions with shallow wells. For the coupled system, in Fig.~\ref{fig:gaptime_2dof}\protect\subref{fig:gaptime_coupled_2dof}, we observe the same time shift as in the uncoupled system, but now the gap time distributions are distributed over a longer interval of time compared to the uncoupled system. Comparing the gap time distributions, we see that if we decrease the depth and increase the flatness of the PES, all the sample trajectories leave the reactants region within a shorter interval of time and the gap times are more close to one another as indicated by the higher peak and narrower width around the mode of the distribution.

\begin{figure}[!ht]
    \subfigure[]{\includegraphics[width=0.45\textwidth]{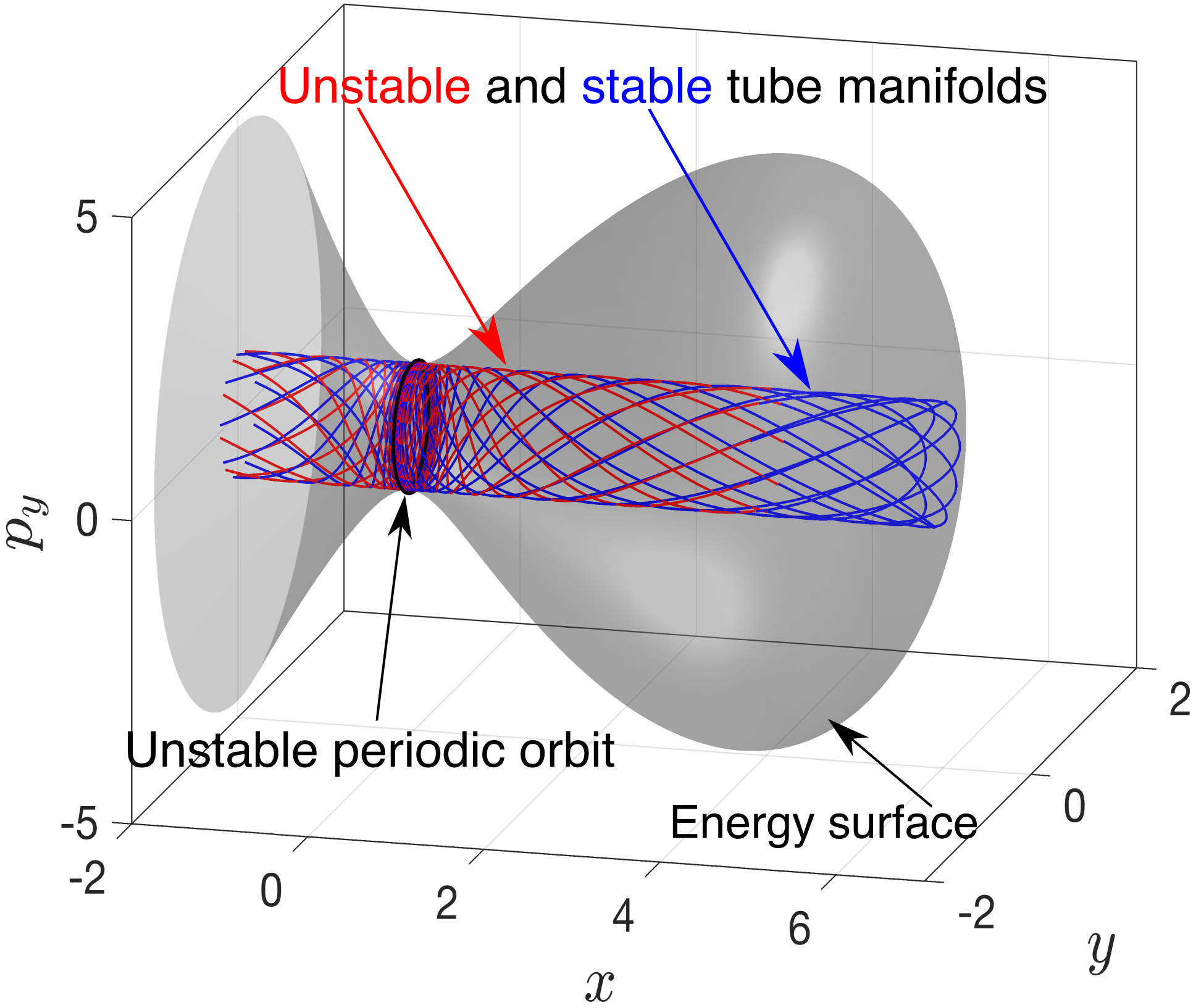}}
    \subfigure[]{\includegraphics[width=0.45\textwidth]{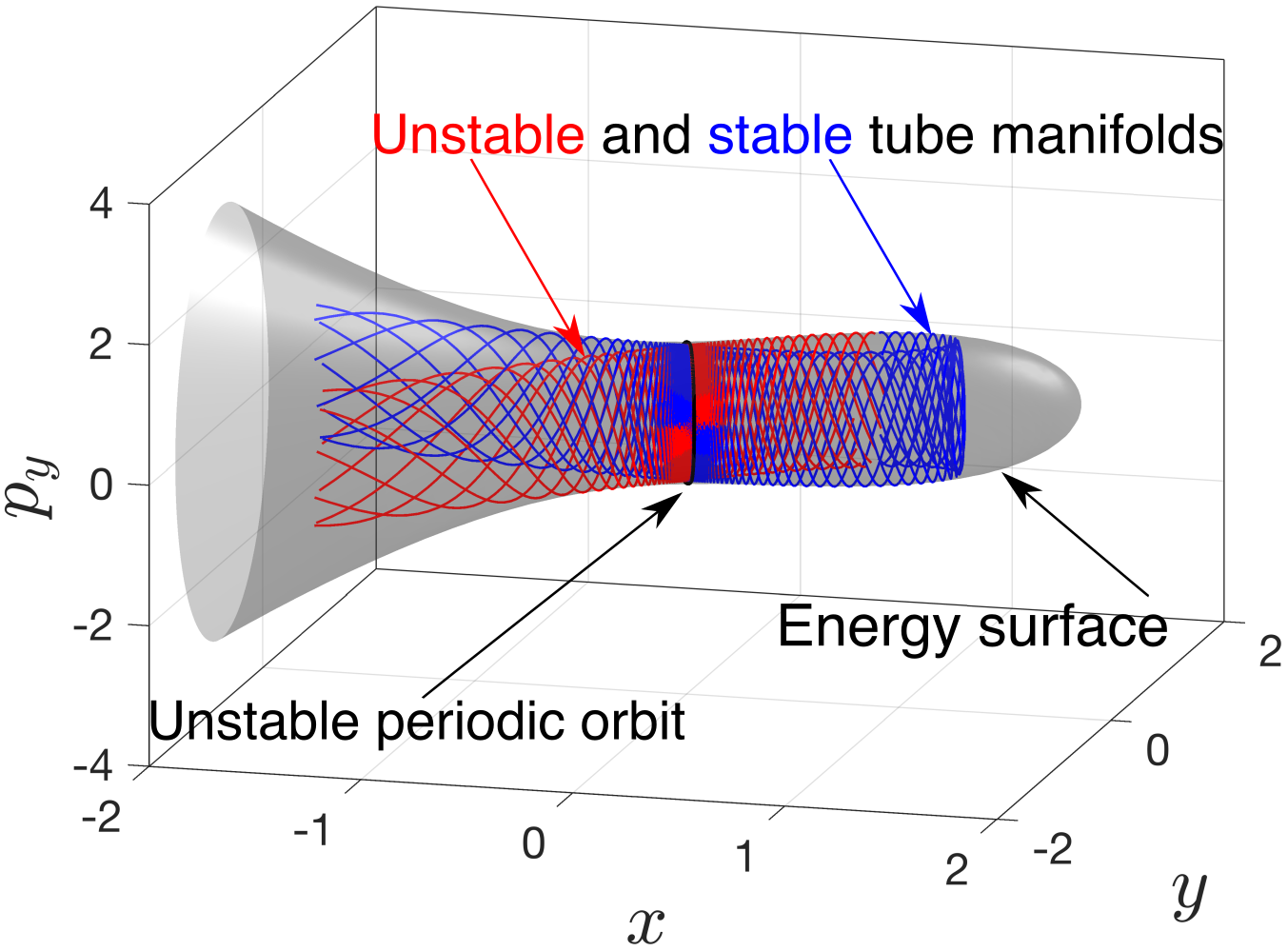}}
    \subfigure[]{\includegraphics[width=0.23\textwidth]{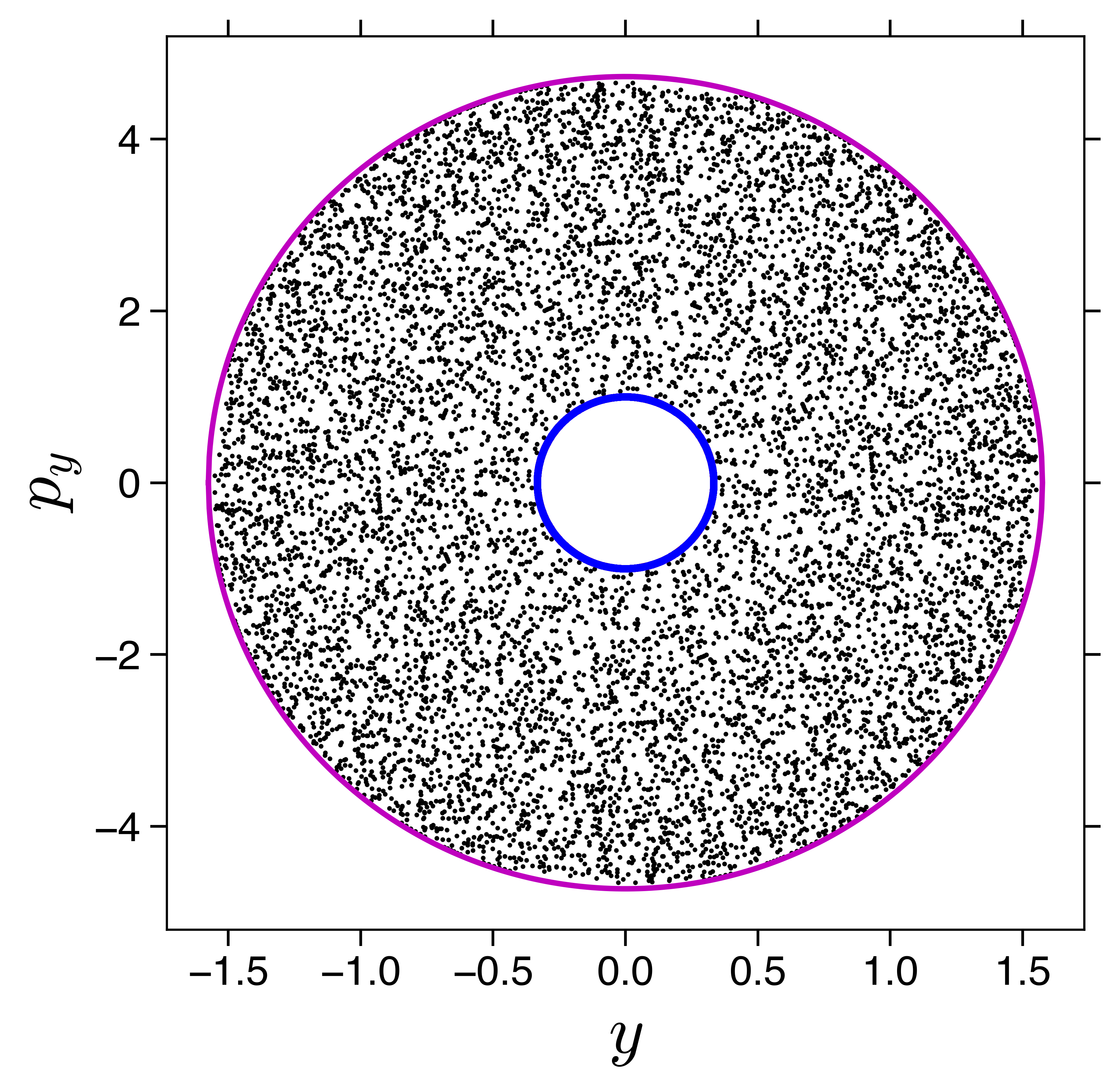}\label{fig:sos_ris_alpha1_epsilon0}}
    \subfigure[]{\includegraphics[width=0.23\textwidth]{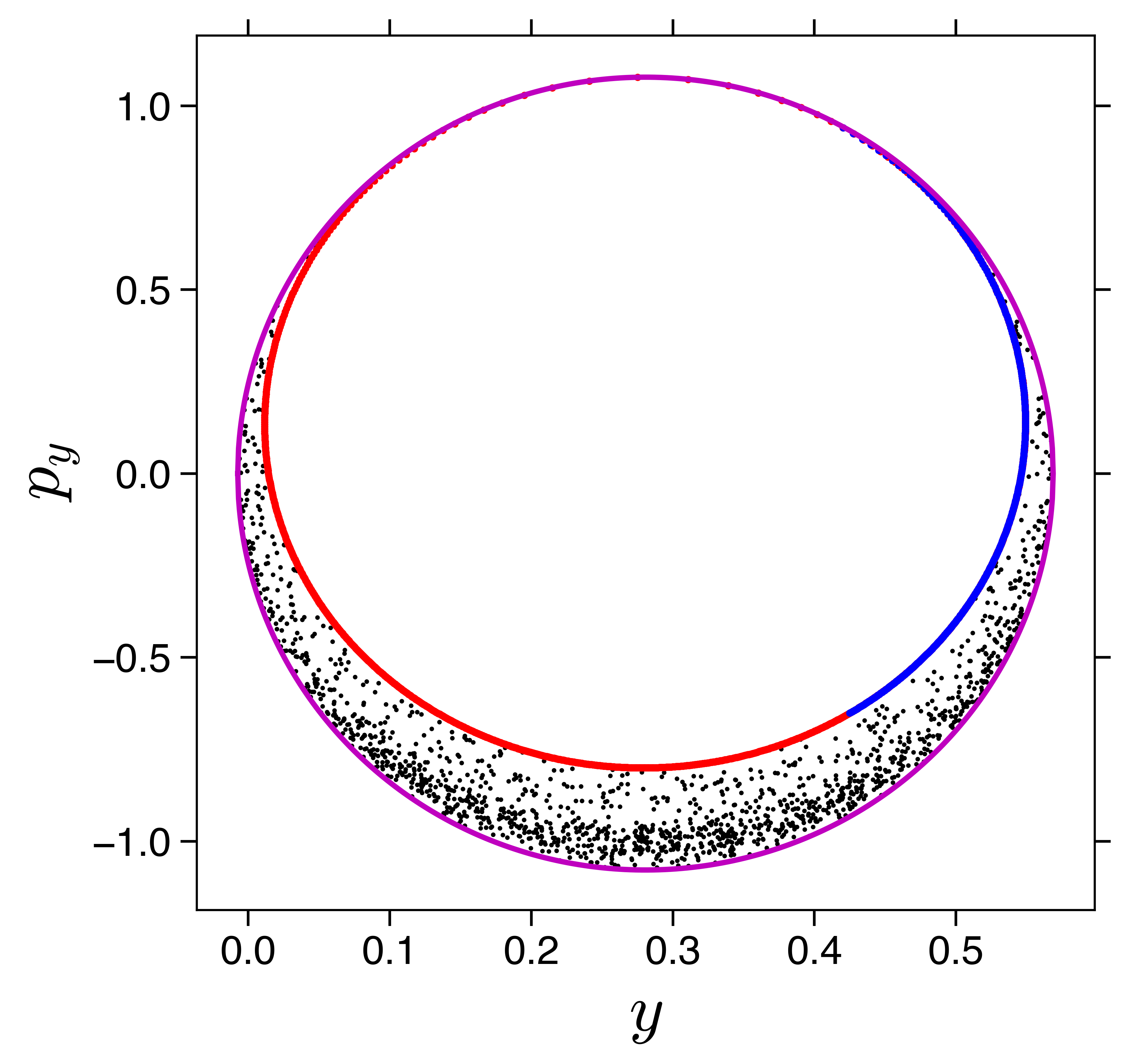}\label{fig:sos_ris_alpha1_epsilon5}}
	\caption{\textbf{Tube (cylindrical) manifolds and surface of section.} Shows the Poincar\'e sections and the reactive island of imminent reaction at total energy $e = 0.5$ for the uncoupled system~\protect\subref{fig:sos_ris_alpha1_epsilon0} $\varepsilon = 0$ and for the coupled system~\protect\subref{fig:sos_ris_alpha1_epsilon5} $\varepsilon = 5$. The intersection of the energy surface with the surface of section~\eqref{eqn:sos_ypy} is shown in magenta. Other system parameters $\mu = 4,\omega = 3, \alpha = 1$ are fixed.}
	\label{fig:sos_ris_alpha125_epsilon05}
\end{figure}

To illustrate the role of the geometry of the invariant manifolds in gap time distributions, we briefly discuss the dynamical fate of the trajectories using Poincar{\'e} surface of section and reactive island theory~\cite{de_leon_intramolecular_1981}. We obtain the surface of section of the trajectories at a fixed total energy by starting the initial conditions on the surface:
\begin{equation}
	U_{yp_y} = \left\{ (x, y, p_x, p_y) \, | \, x = x_e \, , \, p_x(x,y,p_y; e) > 0 \right\}
	\label{eqn:sos_ypy}
\end{equation}

where $x_e$ is the $x$-coordinate of the center-center equilibrium point as defined in Eqn.\ref{eqn:cSpace_eqCoords} and $e$ is the total energy. This positive $p_x$ momentum makes this surface suitable to explain trajectories that are sampled for the gap time distributions discussed above. Using numerical continuation and globalization~\cite{garciagarrido_tilting_2019}, we computed the tube (cylindrical) manifolds shown in the Fig.~\ref{fig:sos_ris_alpha125_epsilon05}(a,b) which mediate the transport of the trajectories across the index-1 saddle in the bottlneck of the energy surface. The Poincar\'e sections and the reactive islands~\cite{marston_reactive_1989} of imminent reactions are shown in Fig.~\ref{fig:sos_ris_alpha125_epsilon05} (c,d) for both the uncoupled and coupled system at a fixed total energy, $e = 0.5$. The reactive island of imminent reaction is the first intersection of the stable and unstable manifold of the unstable periodic orbit at total energy $e$ with the surface of section~\eqref{eqn:sos_ypy}. These intersections are shown as blue and red boundary around the empty (white) region in the Poincar{\'e} surface of sections, while the trapped (during the integration time) trajectories in the well intersect the surface at points shown as black dots. For the system parameters used in the Fig.~\ref{fig:sos_ris_alpha125_epsilon05}, trapped trajectories exhibit chaotic dynamics. 

\subsection{Directional flux}

In this subsection, we discuss the influence of the depth and flatness on the directional flux through the phase space dividing surface (DS) in the two DOF system. It is to be noted that the directional flux calculation for the one DOF system is not valid due to the geometry of the DS.


\subsubsection{Two degree-of-freedom Hamiltonian} 

The directional flux through the phase space DS~\cite{waalkens2004direct,Waalkens2005} is given by the action of the normally hyperbolic invariant manifold (NHIM) which is of dimension $\mathbb{S}^{N - 3}$ in a $N-$dimensional phase space. In the two DOF system, the NHIM is an unstable periodic orbit (UPO) and the directional flux given by the action simplifies to the line integral
\begin{equation}
Q = \int_{\rm{UPO}} \mathbf{p} \cdot d\mathbf{q}    
\end{equation}

For the uncoupled system, the DS is defined by $x = 0$ in the three dimensional energy surface, $H(x,y,p_x,p_y) = e$. The forward~\eqref{eqn:forwds_sn2dof_uncoupled} and backward DS~\eqref{eqn:backds_sn2dof_uncoupled} meet at $p_x=0$ along the UPO defined by 
\begin{align}
\text{ UPO:} \; \left\{ (x,y,p_x,p_y) \in \mathbb{R}^4 \, | \, x = 0, p_x = 0, \dfrac{1}{2} p_y^2 +\dfrac{\omega^2}{2} y^2 = e \right\} 
\end{align}
Thus, for the uncoupled system, the directional flux, $Q$, is given by
\begin{equation}
\begin{aligned}   
Q & = \int_{\text{UPO}} \mathbf{p} \cdot d\mathbf{q} = \int_{0}^{T} \mathbf{p} \cdot \dfrac{d\mathbf{q}}{dt} \ dt \\
Q & = \int_{0}^{T} (p_x \dot{x} + p_y \dot{y} ) \ dt = \int_{0}^{T}  p_y \dot{y} \ dt \\
Q & = \int_{0}^{T}  p_y^2 \ dt = eT = \frac{2 \pi e}{\omega}
\end{aligned}    
\end{equation}
where $T$ is the time period of the UPO and $p_x \dot{x}$ vanishes as $p_x$ coordinate of the UPO is $0$ in the uncoupled system. We calculate the integral by expressing $p_y$ in terms of $t$, which can be done for the uncoupled system using Hamilton's equations of motion~\eqref{hameq_2dof}. We note that this expression is the $N = 2$ case for the flux formula in~\citeauthor{waalkens2004direct}~\cite{waalkens2004direct}.

\begin{figure}[!ht]
	\centering
	\subfigure[]{\includegraphics[width=0.45\textwidth]{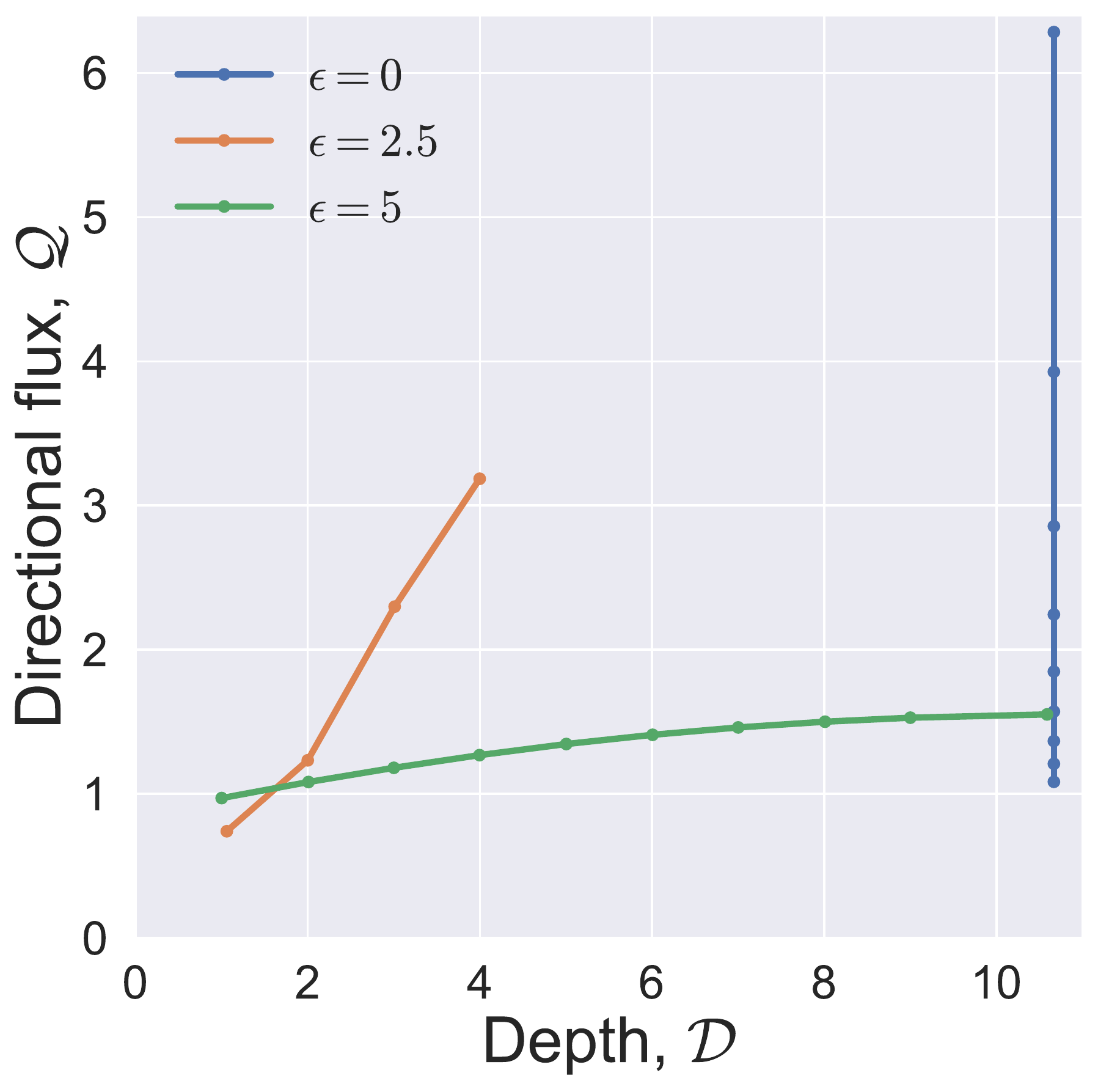}\label{fig:flux_depth_change_omega_2dof}}
	\subfigure[]{\includegraphics[width=0.45\textwidth]{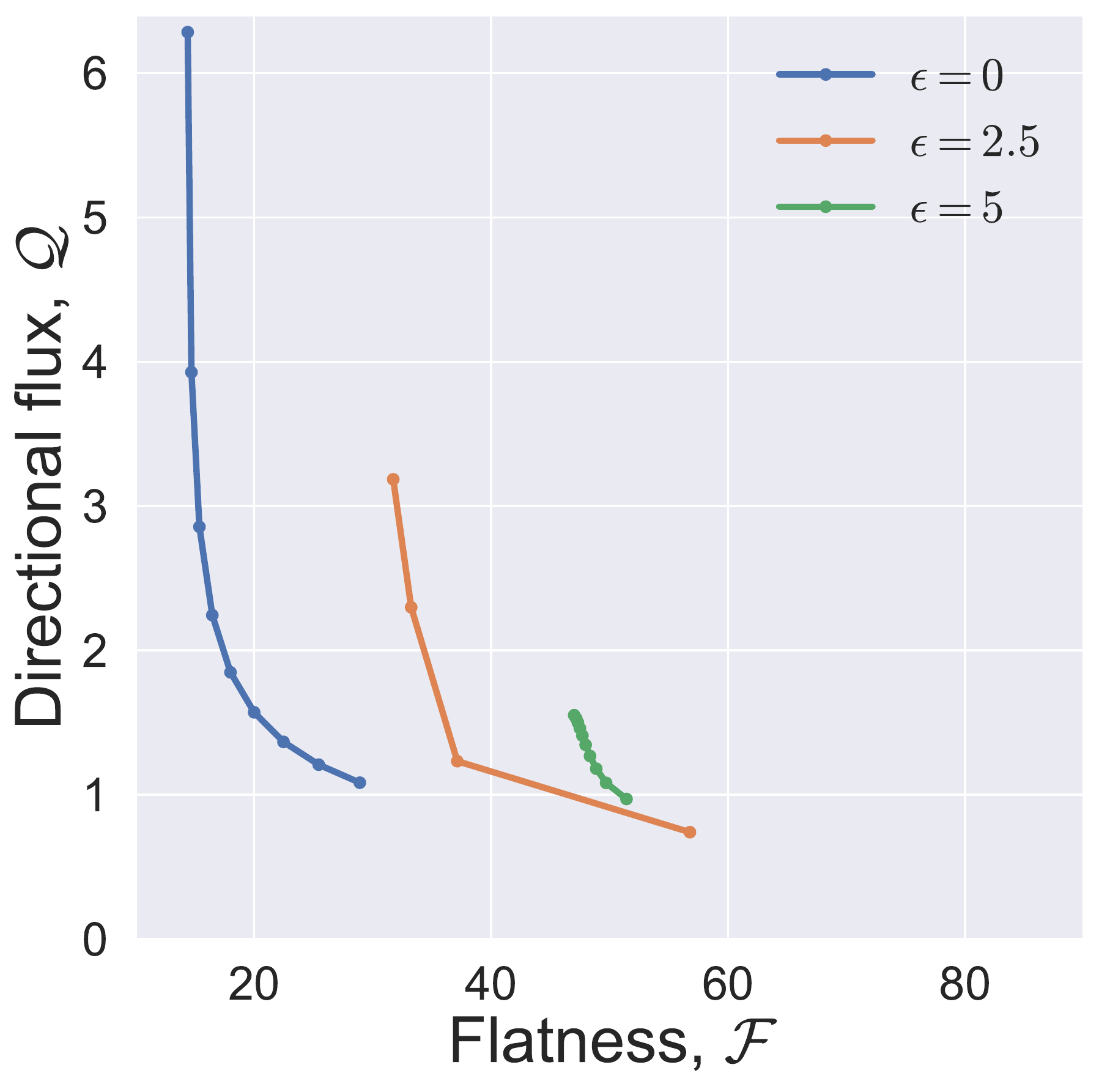}\label{fig:flux_flat_change_omega_2dof}}
	\caption{\textbf{Directional Flux $Q$ for the two DOF sytem.} Shows the reaction flux through the phase space DS increases with increase in \protect\subref{fig:flux_depth_change_omega_2dof} depth and decreases with increase in \protect\subref{fig:flux_flat_change_omega_2dof} flatness of the PES. The blue, orange and green dots correspond to values of directional flux for $\varepsilon=0.0,2.5,5$, respectively. $\mu=4,\alpha=1, e=0.5$ are fixed in all the cases.}
	\label{fig:flux_change_omega_2dof}
\end{figure}

For the coupled system, the directional flux through the DS is given by
\begin{align}   
Q & = \int_{\text{NHIM}} \mathbf{p} \cdot d\mathbf{q} = \int_{0}^{T} \mathbf{p} \dfrac{d\mathbf{q}}{dt} \ dt \notag \\
Q & = \int_{0}^{T} (p_x \dot{x} + p_y \dot{y} ) \ dt = \int_{0}^{T} (p_x \dot{x} + p_y \dot{y} ) \ dt \notag \\
Q & =\int_{0}^{T} p_x^2 + p_y^2 \ dt 
\end{align}    
We evaluate the integral using numerical methods and the unstable periodic orbit is computed using the open source python package~\cite{Lyu2020}. We choose the paramters such that for different $\varepsilon$ values, our $\mathcal{D}$ are roughly integer values between $(0,10]$. We can then calculate $\mathcal{F}$ using the same parameter values. We present the changes in the directional flux for different depth and flatness of the PES in Fig.~\ref{fig:flux_change_omega_2dof}. The depth and flatness of the PES is varied using the parameter $\omega$ while all other system parameters and the total energy are fixed.

We observe that for the uncoupled system in Fig.~\ref{fig:flux_depth_change_omega_2dof}, $\mathcal{D}$ stays constant and this is because our formula for the depth in the uncoupled system does not depend on the parameter $\omega$. However, in Fig.~\ref{fig:flux_flat_change_omega_2dof}, $Q$ decreases as we increase the flatness $\mathcal{F}$ for the uncoupled system, $\varepsilon = 0$. For the coupled system given by $\varepsilon > 0$ in Fig.~\ref{fig:flux_depth_change_omega_2dof}, when we decrease the depth of the PES, the directional flux through the DS also decreases. In Fig.~\ref{fig:flux_flat_change_omega_2dof}, we observe that when the flatness of the PES increases, the directional flux through the phase space DS decreases. In the coupled system, the rate of decay of the directional flux is influenced by the coupling strength, $\varepsilon$ as we observe linear and polynomial behavior for the two values of $\varepsilon$ considered here in Fig.~\ref{fig:flux_change_omega_2dof}. 



\section{Conclusions and outlook} 

In this article, we have presented a formulation for the depth and flatness of a potential energy surface and connected them with quantitative measures of a reaction, such as the reaction probability, directional flux, and gap times. This is done using the one and two degrees of freedom saddle-node Hamiltonian as a preliminary step in understanding a chemical phenomena such as the role of depth and flatness. We observe that decreasing the depth or increasing the flatness of a PES increases the reaction probability in all the systems considered here (cf. Figs.~\ref{fig:samp_pos_reactprob_1dof}(b),~\ref{fig:reactprob_2dof}(a), and ~\ref{fig:reactprob_2dof}(b)). Our investigation also showed that the form of coupling chosen in the two degrees of freedom saddle-node Hamiltonian increases the reaction probability (both the magnitude and rate of growth, cf. Fig.~\ref{fig:reactprob_2dof_ep0} and Fig.~\ref{fig:reactprob_2dof_ep5}) and decreases the peak values of the gap time distribution (cf. Fig.~\ref{fig:gaptime_uncoupled_2dof} and Fig.~\ref{fig:gaptime_coupled_2dof}). In addition, the gap times in the uncoupled system are spread over a small interval ($4 < t < 11$), while for the coupled system gap times are distributed over a longer interval ($0 < t < 20$). It is to be noted that the formulation (sect:~\ref{ssect:depth_flatness_formulas}) presented here is valid for systems with more than two degrees of freedom where the phase space is more than four-dimensional.

In this study, the relationship between the two geometric characteristics of a PES, that is the depth and flatness, and the quantitative measures of a reaction is in agreement with the qualitative understanding from a chemical standpoint. One can use this understanding of the influence of the depth and flatness to assist in predicting reaction rates by incorporating the geometric characteristics of the PES as attributes in a machine-learning model. Related future work would be to use the depth and flatness formulation presented here for studying a PES with multiple bottlenecks and wells. Furthermore, the influence of flatness on roaming and dynamical matching~\cite{agaoglou_chemical_2019,katsanikas_dynamical_2020} that plays a role in product ratio needs to investigated from a quantitative standpoint.

\section*{Author's contributions}

All authors contributed equally to this work.

\section*{Acknowledgements}

We acknowledge the support of  EPSRC Grant No. EP/P021123/1 and Office of Naval Research Grant No. N00014-01-1-0769. The authors would like to acknowledge the London Mathematical Society and School of Mathematics at the University of Bristol for supporting the undergraduate summer research bursary.

\section*{Data availability}

\section*{Data availability}

The code developed for the computations presented in this study are available as open source repository~\cite{saddlenode_code}.

\bibliography{reaction-dynamics}
\bibliographystyle{aipnum4-2.bst}

\appendix
\begin{center}
	\textbf{\LARGE{Appendices}}
\end{center}

\section{Derivation of normal form Hamiltonian for the saddle-node bifurcation}

The following is concerned with the classical Hamiltonian saddle-node bifurcation and using it to model some phenomena of interest relevant to chemical reactions \textemdash~bifurcations of NHIMs, depth and flatness of the potential energy surface. 

\subsection{The ``Standard'' Hamiltonian Saddle-Node Bifurcation}

The  normal form for the one DOF Hamiltonian saddle node bifurcation is given by:
\begin{equation}
H(q, p) = \frac{p^2}{2} - \mu q + \frac{q^3}{3}, 
\end{equation}
with correcponding Hamilton's equations
\begin{eqnarray}
\dot{q} & = & p, \nonumber \\
\dot{p} & = & - q^2 + \mu,
\label{eq:hameq1}
\end{eqnarray}
and $\mu$ is the bifurcation parameter. The equilibria are given by $q^2 = \mu, \: p=0$.

The Jacobian of the vector field is:
\begin{equation}
\left(
\begin{array}{cc}
0 & 1 \\
-2 q & 0
\end{array}
\right).
\end{equation}
We evaluate the Jacobian at each equilibrium point to determine stability:
\begin{equation}
(-\sqrt{\mu}, 0): \qquad \left(
\begin{array}{cc}
0 & 1 \\
2 \sqrt{\mu} & 0
\end{array}
\right),
\qquad
\mbox{saddle},
\end{equation}
\begin{equation}
(\sqrt{\mu}, 0): \qquad \left(
\begin{array}{cc}
0 & 1 \\
-2 \sqrt{\mu} & 0
\end{array}
\right),
\qquad
\mbox{center}.
\end{equation}

\subsection{Fixing the Saddle Point at the Origin}

In the Hamiltonian saddle node described above the equilibria move as $\mu$ is varied. It will be useful to fixe the saddle point at the origin. To do this we introduce the following coordinate transformation: $q=x-\sqrt{\mu}, \; p=y$. Substituting this into \eqref{eq:hameq1} gives:
\begin{eqnarray}
\dot{q} =  \dot{x} & = &  p = y, \nonumber \\
\dot{p} = \dot{y} & = & - (x- \sqrt{\mu})^2 + \mu , \nonumber \\
& = & -x^2  + 2x \sqrt{\mu}.
\end{eqnarray}
or
\begin{eqnarray}
\dot{x} & = & y, \nonumber \\
\dot{y } & = & 2  \sqrt{\mu} x -x^2,
\label{eq:hameq2}
\end{eqnarray}
with corresponding Hamiltonian:
\begin{equation}
H(x, y) = \frac{y^2}{2} - \sqrt{\mu} x^2 + \frac{x^3}{3}.
\end{equation}

The equilibria are given by: 
\begin{equation}
(x, y) = (0, 0), \, \, (2 \sqrt{\mu}, 0).
\end{equation}
The Jacobian of the vector field is:
\begin{equation}
\left(
\begin{array}{cc}
0 & 1 \\
-2 x + 2 \sqrt{\mu} & 0
\end{array}
\right).
\end{equation}

We evaluate the Jacobian at each equilibrium point to determine stability:

\[
(0, 0) : \qquad \left(
\begin{array}{cc}
0 & 1 \\
2 \sqrt{\mu} & 0
\end{array}
\right),
\qquad
\mbox{saddle},
\]

\[
(2 \sqrt{\mu}, 0) : \qquad \left(
\begin{array}{cc}
0 & 1 \\
-2 \sqrt{\mu} & 0
\end{array}
\right),
\qquad
\mbox{center}.
\]

\subsection{Parameter to Control the Depth of the Potential Energy Surface} 

The depth of the potential well is controlled by the cubic term in the potential energy surface. Therefore we introduce a parameter that allows us to vary the amplitude of this  term:
\begin{equation}
H(x, y) = \frac{y^2}{2} - \sqrt{\mu} x^2 + \frac{\alpha x^3}{3}, \qquad \alpha >0.
\end{equation}

\begin{eqnarray}
\dot{x} & = &  y, \nonumber \\
\dot{y} & = & 2\sqrt{\mu} x - \alpha x^2.
\end{eqnarray}
The equilibria are given by:
\begin{equation}
(x, y) = (0, 0), \, \, \left(\frac{2}{\alpha} \sqrt{\mu}, 0 \right).
\end{equation}
The Jacobian of the vector field is given by:

\begin{equation}
\left(
\begin{array}{cc}
0 & 1 \\
2 \sqrt{\mu} -2 \alpha x & 0
\end{array}
\right).
\end{equation}
We evaluate the Jacobian at the equilibria to determine their stability:

\[
(0, 0) : \qquad \left(
\begin{array}{cc}
0 & 1 \\
2 \sqrt{\mu} & 0
\end{array}
\right),
\qquad
\mbox{saddle}.
\]

\[
\left(\frac{2}{\alpha} \sqrt{\mu}, 0\right) : \qquad \left(
\begin{array}{cc}
0 & 1 \\
-3 \sqrt{\mu} & 0
\end{array}
\right),
\qquad
\mbox{center}.
\]

``Depth'' of the potential energy surface is determined by the difference between the potential evaluated at the saddle minus the potential evaluated at the center (minumum of the well). The potential energy function is given by:
\begin{equation}
V(x) =  - \sqrt{\mu} x^2 + \frac{\alpha x^3}{3}, \qquad \alpha >0,
\end{equation}
and this difference is given by:
\begin{eqnarray}
V(0) - V\left(\frac{2}{\alpha} \sqrt{\mu}\right) & = & \frac{4}{\alpha^2} \mu \sqrt{\mu} - \frac{\alpha}{3} \frac{4}{\alpha^2} \mu \frac{2}{\alpha} \sqrt{\mu}, \nonumber \\
& = & \left(\frac{4}{\alpha^2} - \frac{8 \alpha}{3 \alpha^3} \right) \mu \sqrt{\mu}, \nonumber \\
& = & \left(\frac{12}{3 \alpha^2} - \frac{8 }{3 \alpha^2} \right) \mu \sqrt{\mu}, \nonumber \\
& = & \frac{4}{3 \alpha^2} \mu \sqrt{\mu}.
\end{eqnarray}

Hence for a fixed $\mu$ (that is, the distance apart of the two equilibria) the potential is made less deep by taking large $\alpha$.

\section{Derivation of the expression for the ratio of the bottleneck-width and well-width for the coupled two DOF system}
The following expression is defined as $A$:
\begin{equation}
A = e - \left(\dfrac{\alpha}{3}\, x^3- \sqrt{\mu} \, x^2  + \dfrac{\varepsilon}{2} x^2 \right) + \frac{2 x^2 \varepsilon^2}{4(\omega^2+\varepsilon)}
\end{equation}

Then the ratio of the bottleneck-width and well-width can be written as

\begin{widetext}
\begin{align}
R_{bw} &= \dfrac{w_b}{w_w} =\dfrac{\text{width of the bottleneck}}{\text{width of the well}} \notag \\
&= \dfrac{\sqrt{\dfrac{2e}{\omega^2+ \varepsilon}}}{\sqrt{A|_{x=x^e,V=e}} \sqrt{\dfrac{2}{\omega^2 + \varepsilon}}} \notag \\
&= \sqrt{\dfrac{e}{e- (\dfrac{\alpha}{3}\, (x^e)^3- \sqrt{\mu} \, (x^e)^2  + \dfrac{\varepsilon}{2} (x^e)^2) + \dfrac{2 (x^e)^2 \varepsilon^2}{4(\omega^2+\varepsilon)}}} \notag \\
&= \sqrt{\dfrac{e}{e- (\dfrac{\alpha}{3}\, x^e- \sqrt{\mu} \,  + \dfrac{\varepsilon}{2} - \dfrac{2  \varepsilon^2}{4(\omega^2+\varepsilon)})((x^e)^2)}} \notag \\
&= \sqrt{\dfrac{e}{e- (\dfrac{\alpha}{3}\, \dfrac{1}{\alpha} (2\sqrt{\mu}- \dfrac{\omega^2 \varepsilon}{\omega^2+\varepsilon})-\sqrt{\mu} \,  + \dfrac{\varepsilon}{2} - \dfrac{2  \varepsilon^2}{4(\omega^2+\varepsilon)} )\dfrac{1}{\alpha^2}(2\sqrt{\mu}- \dfrac{\omega^2 \varepsilon}{\omega^2+\varepsilon})^2}}  \notag \\
&= \sqrt{\dfrac{e}{e- \dfrac{1}{\alpha^2}(\dfrac{2\sqrt{\mu}}{3}\, - \dfrac{\omega^2 \varepsilon}{3(\omega^2+\varepsilon)}-\sqrt{\mu} \,  + \dfrac{\varepsilon}{2} - \dfrac{2  \varepsilon^2}{4(\omega^2+\varepsilon)} )(2\sqrt{\mu}- \dfrac{\omega^2 \varepsilon}{\omega^2+\varepsilon})^2}}  \notag \\
&= \sqrt{\dfrac{e}{e- \dfrac{1}{\alpha^2}(-\dfrac{\sqrt{\mu}}{3} + \dfrac{\omega^2\varepsilon}{6(\omega^2+\varepsilon)} )(2\sqrt{\mu}- \dfrac{\omega^2 \varepsilon}{\omega^2+\varepsilon})^2}}  \notag \\
&= \sqrt{\dfrac{e}{e+ \dfrac{1}{6\alpha^2}(2\sqrt{\mu}- \dfrac{\omega^2 \varepsilon}{\omega^2+\varepsilon})^3}}  \notag  \\
&= \sqrt{\dfrac{e}{e + \mathcal{D_{\varepsilon}}}}
\end{align}
\end{widetext}

\end{document}